\documentclass[12pt,a4paper]{article}
\usepackage{cite,color,graphics,amsmath,rotating}
\usepackage{epstopdf}

\begin{document}

\setlength{\unitlength}{1mm}
\renewcommand{\arraystretch}{1.4}


\def\micro {{\tt micrOMEGAs}}
\def\micromegas {{\tt micrOMEGAs\,2.4}}
\def\suspect {{\tt SuSpect}}

\def\ma{M_A}
\def\ra{\rightarrow}
\def\snr{\tilde{\nu}_R}
\def\lsp{\tilde{\nu}_1}
\def\mlsp{m_{\tilde{\nu}_1}}
\def\msnut{m_{\tilde{\nu}_2}}
\def\snl{\tilde{\nu}_L}
\def\snt{\tilde{\nu}_2}
\def\mstau{m_{\tilde{\tau}}}
\def\mneut{m_{\tilde{\chi}^0_1}}
\def\mchi{m_{\tilde{\chi}^0_i}}
\def\mneutt{m_{\tilde{\chi}^0_2}}
\def\mneuth{m_{\tilde{\chi}^0_3}}
\def\mneutf{m_{\tilde{\chi}^0_4}}
\def\mchar{m_{\tilde{\chi}^\pm_1}}
\def\mchart{m_{\tilde{\chi}^\pm_2}}
\def\msel{m_{\tilde{e}_L}}
\def\mser{m_{\tilde{e}_R}}
\def\mslo{m_{\tilde{\tau}_1}}
\def\mslt{m_{\tilde{\tau}_2}}
\def\msul{m_{\tilde{u}_L}}
\def\msur{m_{\tilde{u}_R}}
\def\msdl{m_{\tilde{d}_L}}
\def\msdr{m_{\tilde{d}_R}}
\def\msto{m_{\tilde{t}_1}}
\def\mstt{m_{\tilde{t}_2}}
\def\msbo{m_{\tilde{b}_1}}
\def\msbt{m_{\tilde{b}_2}}
\def\sw{s_W}
\def\cw{c_W}
\def\bsmu{B_s\rightarrow \mu^+\mu^-}
\def\ca{\cos\alpha}
\def\cb{\cos\beta}
\def\sa{\sin\alpha}
\def\sb{\sin\beta}
\def\tb{\tan\beta}
\def\si{\sigma^{\rm SI}}
\def\sip{\sigma^{\rm SI}_{\lsp p}}
\def\ssi{\sigma^{\rm SI}_{\lsp N}}
\def\ssd{\sigma^{\rm SD}_{\chi N}}
\def\sd{\sigma^{\rm SD}}
\def\sdp{\sigma^{\rm SD}_{\chi p}}
\def\sdn{\sigma^{\rm SD}_{\chi n}}
\def\msl{M_{\tilde l}}
\def\msq{M_{\tilde q}}
\def\Omg{\Omega h^2}
\def\sip{\sigma^{SI}_{\chi p}}
\def\amu{\delta a_\mu}
\def\neuto{\tilde\chi^0_1}
\def\neuti{\tilde\chi^0_i}
\def\neutt{\tilde\chi^0_2}
\def\neuth{\tilde\chi^0_3}
\def\neutf{\tilde\chi^0_4}
\def\chargi{\tilde\chi^\pm_i}
\def\charg{\tilde\chi^\pm_1}
\def\chargt{\tilde\chi^\pm_2}
\def\gluino{\tilde{g}}
\def\ul{\tilde{u}_L}
\def\ur{\tilde{u}_R}
\def\stau{\tilde{\tau}}
\def\sl{\tilde{l}}
\def\sq{\tilde{q}}
\def\bone{B^1}
\def\sneutrino{\tilde\nu}
\def\msnu{m_{\tilde\nu_1}}
\def\msnut{m_{\tilde\nu_2}}
\def\anu{A_{\tilde\nu}}
\def\Anu{A_{\tilde\nu}}
\def\sn{\sin\theta_{\tilde\nu}}
\def\cn{\cos\theta_{\tilde\nu}}
\def\tsn{\theta_{\tilde\nu}}
\def\pbar{\bar{p}}
\newcommand\lsim{\mathrel{\rlap{\lower4pt\hbox{\hskip1pt$\sim$}}\raise1pt\hbox{$<$}}}
\newcommand\gsim{\mathrel{\rlap{\lower4pt\hbox{\hskip1pt$\sim$}}\raise1pt\hbox{$>$}}}

\def\note{\marginpar{\bf !!}}
\def\todo{\marginpar{\bf todo}}
\def\verify{\marginpar{\bf verify}}


\vspace*{4mm}

\begin{center}

{\Large\bf Light mixed sneutrinos as thermal dark matter} \\[8mm]

{\large   G.~B\'elanger$^1$, M.~Kakizaki$^1$, S.~Kraml$^2$, E.~K.~Park$^{1,3}$ 
and A.~Pukhov$^4$}\\[4mm]
{\it 1) LAPTH, Univ. de Savoie, CNRS, B.P.\,110, F-74941 Annecy-le-Vieux, France\\
     2) LPSC, UJF Grenoble 1, CNRS/IN2P3, 53 Avenue des Martyrs,\\ F-38026 Grenoble, France\\
     3) Physikalisches Institut, Universit\"at Bonn, Nussallee 12, D-53115 Bonn, Germany\\
     4) Skobeltsyn Inst. of Nuclear Physics, Moscow State Univ., Moscow 119992, Russia 
}\\[2mm]

\end{center}

\begin{abstract}
In supersymmetric models with Dirac neutrino masses, a left-right mixed sneutrino 
can be a viable dark matter candidate. We examine the MSSM+$\tilde\nu_R$ parameter 
space where this is the case with particular emphasis on light sneutrinos with masses below 
10 GeV. We discuss implications for direct and indirect dark matter searches, including 
the relevant uncertainties, as well as consequences for collider phenomenology.
\end{abstract}

\vspace*{4mm}

\section{Introduction}

A simple extension of the Standard Model (SM) by right-handed neutrinos provides 
the framework for describing neutrino masses and the observed neutrino 
oscillations~\cite{Pontecorvo:1967fh,Gribov:1968kq} (for reviews 
see, e.g., \cite{Bilenky:1998dt,GonzalezGarcia:2007ib}).
Current observations, however, do not allow to establish the Majorana or 
Dirac nature of neutrinos. While the smallness of the neutrino mass can be 
naturally explained by introducing Majorana mass terms and making use of 
the see-saw mechanism, Dirac masses for neutrinos with very small Yukawa 
couplings are a viable and interesting alternative. 
In supersymmetric models, one may naturally obtain very light Dirac neutrino 
masses from F-term SUSY breaking~\cite{ArkaniHamed:2000bq}. In addition to 
providing an explanation for neutrino masses, this class of supersymmetric 
models offers an interesting alternative dark matter (DM) candidate, the 
sneutrino. 
Indeed in these models one can generate a weak-scale trilinear $A_{\tilde\nu}$ 
term that is not proportional to the small neutrino Yukawa couplings. 
Thus large mixing between left-handed (LH) and right-handed (RH) 
sneutrinos can be induced even though the 
Yukawa couplings are extremely small.
This is in sharp contrast with the usual MSSM where the trilinear $A$ terms 
are proportional to the Yukawa couplings so that  mixing effects can be 
neglected for the first two generations of sfermions. 

The lightest sneutrino can thus become the lightest SUSY particle (LSP) and 
a viable thermal DM candidate. 
Because of the large sneutrino mixing, the mainly RH sneutrino is no longer 
sterile, its couplings to SM gauge and Higgs bosons are driven by the mixing 
with its LH partner. Sufficient mixing provides efficient annihilation so that 
one can obtain a value for the relic density of $\Omega h^2 \simeq 0.11$ as 
extracted from cosmological observations~\cite{Dunkley:2008ie,Komatsu:2010fb,Jarosik:2010iu}. 

Direct detection (DD) experiments pose severe constraints on Dirac or complex 
scalar, i.e.\ not self-conjugated, DM particles because the spin-independent 
elastic scattering cross-section receives an important contribution from Z 
exchange, which typically exceeds experimental bounds. 
In the mixed sneutrino model, this cross-section is suppressed by the sneutrino 
mixing angle. Therefore, on the one hand a viable sneutrino DM candidate requires 
enough mixing to provide sufficient pair-annihilation, on the other hand the 
mixing should not be too large in order not to exceed the DD limits. 
Here we will explore the parameter space of the model where these conditions 
are satisfied.

It is intriguing that a mixed sneutrino also opens the possibility for a 
supersymmetric DM candidate below 10~GeV.  
Light DM candidates have received a lot of attention recently~\cite{Feng:2008qn,Kim:2009ke,
Cerdeno:2009dv,Fitzpatrick:2010em,Andreas:2010dz,Essig:2010ye,Chang:2010yk,Bae:2010hr,
Mambrini:2010dq,Das:2010ww,Lavalle:2010yw}  
because of results of DD experiments that show hints of events compatible 
with light DM. 
This includes the modulation signal from DAMA~\cite{Bernabei:2008yi} 
as well as recent results from CoGeNT~\cite{Aalseth:2010vx} and 
CDMS~\cite{Ahmed:2009zw}. 
The best fit values for the mass and the cross-section do not overlap when these
results are interpreted as a spin-independent contribution, nevertheless analyses 
taking into account the signal's dependence on the DM velocity distribution 
have shown \cite{Kopp:2009qt,Fitzpatrick:2010em,Savage:2010tg} that the observed events can be 
compatible with the null results obtained by other experiments such as 
Xenon~\cite{Angle:2007uj,Aprile:2010um}.  
In addition, CRESST-II has very recently reported 32 events with an expected 
background of $8.7\pm 1.4$, compatible with a DM mass of 15~GeV or below 
and a spin-independent cross-section of a few times $10^{-5}$~pb \cite{cresst}. 

Whether or not these events are confirmed, the possibility of light DM with large 
elastic scattering cross-sections remains interesting and particularly challenging 
to probe because experiments suffer from a severe loss of sensitivity at low masses. 
Furthermore, the mixed sneutrino revives the possibility of light DM in the MSSM:  
although it is possible to find a light neutralino that has the right properties 
to satisfy DM constraints in the MSSM with non-universal gaugino masses, this 
scenario is incompatible with additional constraints on the model~\cite{Belanger:2003wb,Feldman:2010ke}. 
Indeed efficient annihilation requires additional light particles, for example a 
second doublet Higgs with mass in the 100~GeV range, which is strongly constrained 
by B-physics processes, in particular $\bsmu$~\cite{Aaltonen:2007kv}. 
A light neutralino is still possible in singlet extensions of the MSSM where  
new Higgs singlets provide additional possibilities for efficient annihilation of the 
lightest neutralino~\cite{Das:2010ww,Vasquez:2010ru}.

The phenomenology of the mixed-sneutrino model that we examine here was 
first investigated in~\cite{ArkaniHamed:2000bq}. Indirect detection signatures  
were discussed in~\cite{Arina:2007tm}, and LHC signatures in~\cite{Thomas:2007bu}.\footnote{Many  
more studies of sneutrino DM have been performed in other models, like models with extra singlets or 
models with Majorana neutrino masses, see e.g., \cite{Hall:1997ah,Kolb:1999wx,Asaka:2005cn, 
Asaka:2006fs,Lee:2007mt,Arina:2008yh,Arina:2008bb,Cerdeno:2008ep,Deppisch:2008bp,Allahverdi:2009ae,
Cerdeno:2009dv,Demir:2009kc,Allahverdi:2009se,Cerdeno:2009zz,Allahverdi:2009kr,
Kumar:2009sf,MarchRussell:2009aq}.}
We extend on these analyses in several ways.
First of all, in contrast to the above mentioned studies, we here concentrate on 
light DM with mass of about 10~GeV and below. 
Second, we explore the parameter space of the model that gives a consistent sneutrino 
DM candidate using up-to-date constraints on elastic scattering cross-sections, 
examining also the effects of uncertainties in, e.g., the DM velocity distribution. 
In our scans, we take into account radiative corrections to the SUSY and Higgs spectrum;  
in particular we include the 1-loop corrections to the sneutrino mass originating from the Higgs contribution, 
and those to the light Higgs mass originating from the large $\Anu$ term. 
Moreover, we consider both the case of one and of three sneutrino flavours,
assuming complete degeneracy in the three flavour case.  
For the allowed scenarios, we explore the consequences for DD as well as for 
indirect detection in photons, antiparticles and neutrinos. 
Finally, we explore the consequences for searches at the LHC and ILC. 

We characterize the scenarios that satisfy the DD constraints, including those that 
are within the region favoured by CoGeNT (and maybe also CRESST). 
The allowed scenarios have specific characteristics which include, e.g.,  
dominantly invisible Higgs decays. Besides, if the charged sleptons are heavier 
than the $\tilde\chi^0_2$ and $\tilde\chi^\pm_1$, as is the case over most 
of the valid parameter space, this implies dominantly invisible decays of 
neutralinos ($\tilde\chi^0_{1,2}\to\nu\tilde\nu_1$) and single-lepton decays of charginos 
($\tilde\chi^\pm_1\to \ell^\pm\tilde\nu_1$).  

The paper is organized as follows. 
Section~\ref{sec:model} describes the framework of our analysis,  
giving details on the model, the mass spectrum, and radiative corrections.  
Section~\ref{sec:constraints} then discusses collider constraints, sneutrino annihilation 
and direct detection. 
The relic density and DD predictions for the one light sneutrino  
case are analyzed in Section~\ref{sec:onesnu}, including a discussion 
of astrophysical uncertainties.  
Results for three degenerate light sneutrinos are presented in 
Section~\ref{sec:threesnu}. 
Signatures in indirect detection are discussed in Section~\ref{sec:indirect} 
and collider signatures in Section~\ref{sec:colliders}. 
A summary and conclusions are given in Section~\ref{sec:conclusions}.
The Appendix contains Feynman rules for the relevant sneutrino interactions.

All numerical results have been obtained with  \micro~\cite{Belanger:2008sj,Belanger:2010gh}, 
linked to an appropriately modified version of \suspect~\cite{suspect}.

\section{Framework}\label{sec:model}

\subsection{Mixed sneutrinos}\label{sec:model1}

The framework for our study is the model of \cite{ArkaniHamed:2000bq} with only Dirac masses 
for sneutrinos. In this case, the usual MSSM soft-breaking terms are extended by
\begin{equation}
  \Delta {\cal L}_{\rm soft} = m^2_{\tilde N_i}  |\tilde N_i |^2 +  
                                            A_{\tilde\nu_i} \tilde L_i \tilde N_i H_u + {\rm h.c.} \,,
\end{equation}
where ${m}^2_{\tilde{N}}$ and $A_{\tilde\nu}$ are weak-scale soft terms, which we assume to 
be flavour-diagonal. Note that the lepton-number violating bilinear term, which appears 
in case of Majorana neutrino masses, is absent. 
Neglecting the tiny Dirac masses, the $2\times2$ sneutrino mass matrix for one generation is 
given by 
\begin{equation}
 {\cal M}^2_{\tilde\nu} =
  \left( \begin{array}{cc}
   {m}^2_{\widetilde{L}} +\frac{1}{2} m^2_Z \cos 2\beta  &  \frac{1}{\sqrt{2}} A_{\tilde\nu}\, v \sin\beta\\
   \frac{1}{\sqrt{2}}    A_{\tilde\nu}\, v \sin\beta&  {m}^2_{\widetilde{N}}
  \end{array}\right) \,.
\label{eq:sneutrino_tree}
\end{equation}
Here ${m}^2_{\tilde{L}}$ is the SU(2) slepton soft term, $v^2=v_1^2+v_2^2=(246\;{\rm GeV})^2$ 
with $v_{1,2}$ the Higgs vaccuum expectation values, and $\tan\beta=v_2/v_1$.  
The main feature of this model is that the ${m}^2_{\widetilde{L}}$, ${m}^2_{\widetilde{N}}$
and $A_{\tilde\nu}$ are all of the order of the weak scale, and 
$A_{\tilde\nu}$ does not suffer any suppression from Yukawa couplings. 
In the following, we will always assume $m_{\tilde N}<m_{\tilde L}$ so that the lighter mass 
eigenstate, $\tilde\nu_1$, is mostly a $\tilde\nu_R$. 
It is in fact quite natural to obtain this relation when embedding the model 
in a GUT scale model, because the renormalization group
running of $m_{\tilde L}$ is governed by $M_2$, while for $m_{\tilde N}$ the
running at 1~loop is driven exclusively by the $A_{\tilde\nu}$ term, since 
$\tilde\nu_R$ is a SM singlet. 
 
A large $A_{\tilde\nu}$ term in the sneutrino mass matrix will induce a significant 
mixing between the LH and RH states, 
\begin{align}
  \tilde\nu_1 &= \cos\theta_{\tilde\nu} \, \tilde\nu_R - \sin\theta_{\tilde\nu}\,  \tilde\nu_L \,, \\
  \tilde\nu_2 &= \sin\theta_{\tilde\nu} \, \tilde\nu_R + \cos\theta_{\tilde\nu}\,  \tilde\nu_L \,, 
\end{align}
where $m_{\tilde\nu_1} < m_{\tilde\nu_2}$ and the mixing angle 
\begin{equation}
   \theta_{\tilde\nu} = \frac{1}{2}\, \sin^{-1}\!\left( 
    \frac{\sqrt{2}\anu\, v\sin\beta }{ {m}^2_{\tilde\nu_2}-{m}^2_{\tilde\nu_1} } \right) \,.
\end{equation}
Notice that for fixed $\sn$, $\anu$ is proportional to $m_{\tilde{\nu}_2}^2-m_{\tilde{\nu}_1}^2$.
This means that $\anu$ is of the same order as other soft terms in the sneutrino sector.

A large value of $\anu$ can induce a large splitting between the two mass eigenstates 
even if ${m}^2_{\tilde{L}}$ and ${m}^2_{\tilde{N}}$ are of the same order, leading 
to scenarios where $m_{\tilde\nu_1} \ll m_{\tilde\nu_2},m_{\tilde{l}_L}$. In this way, 
$\lsp$ can naturally be driven much below the neutralino masses. The model can easily 
be generalized to three generations. When doing so we will neglect for simplicity 
any flavour mixing in the sneutrino sector. 

The couplings of the mostly sterile $\tilde\nu_1$ are those of the LH sneutrino 
suppressed by a factor $\sin^2\theta_{\tilde\nu}$ due to mixing. In addition, there 
is a new direct coupling between the Higgs bosons and the LH and RH sneutrino
components proportional to $\anu$. 
The couplings to the Z and light Higgs boson will play a crucial role both for  
annihilation processes and for the elastic scattering cross-section. 
The coupling of sneutrinos to neutrinos and neutralinos is dominated by the 
wino component of the neutralinos and will be important for the
annihilation of sneutrinos into neutrinos. 
The relevant Feynman rules are given explicitly in the Appendix.

\subsection{Particle spectrum}\label{sec:spectrum}

We assume a model with soft terms defined at the weak scale  
and unification of gaugino masses at the GUT scale. The latter leads to $M_2 \simeq 2M_1 \simeq M_3/3$
at the weak scale. For the sneutrino sector, we take the masses and mixing angle, 
$m_{\lsp}$, $m_{\tilde \nu_2}$ and $\theta_{\tilde\nu}$, as input parameters, and 
compute the ${m}_{\widetilde{L}}$, ${m}_{\widetilde{N}}$ and $A_{\tilde\nu}$. 
This also fixes the corresponding LH charged slepton mass term; for the RH one we 
assume $m_{\widetilde R}=m_{\widetilde L}$. 
Note that this choice has no effect as concerns DM properties,  
but can have implications for collider searches as will be discussed in Section~\ref{sec:colliders}.

In the one-generation case, we assume that only the tau-sneutrino is light and all others are heavy, 
with soft masses of 1~TeV. In the three-generation case, on the other hand, complete degeneracy 
between the slepton generations is assumed.  

For the squark sector, we assume a common soft mass $m_{\tilde q}=1$~TeV and take 
$A_t=-1$~TeV in order to avoid the constraint on the light Higgs mass. 
Other trilinear couplings for charged sparticles are neglected. 
The higgsino mass, pseudoscalar Higgs mass and $\tan\beta$ are also input parameters, 
we fix them to $\mu=800$~GeV, $M_A=1$~TeV and $\tan\beta=10$. 

We use a modified version of \suspect\ \cite{suspect} for the spectrum calculation.
The original \suspect\  includes the 1-loop radiative corrections to neutralino, chargino and squark 
masses; corrections to Higgs masses are implemented at the two-loop level. We have extended 
it to include RH sneutrinos, and implemented 1-loop radiative corrections to sneutrino 
masses as well as those to the light Higgs mass induced by the $\anu$ term.  

\subsubsection{Radiative corrections to sneutrino masses}

\begin{figure}[t!]\centering
\includegraphics[width=0.6\textwidth]{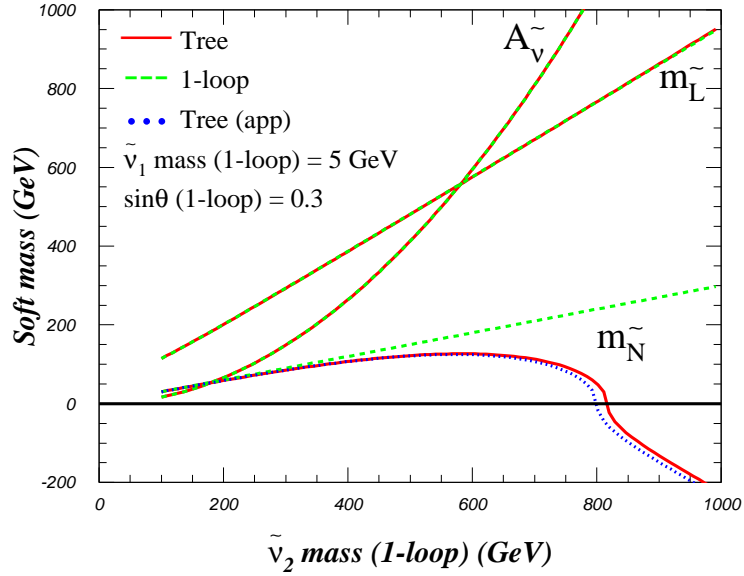}
\vspace*{-2mm}
\caption{Tree-level and 1-loop soft terms $m_{\widetilde{L}}$, $m_{\widetilde{N}}={\rm sign}(m^2_{\widetilde{N}}) |m^2_{\widetilde{N}}|^{1/2}$  
          and $A_{\tilde\nu}$ that give $m_{\tilde{\nu}_1}|_{\rm 1-loop} = 5$~GeV and 
          $\sin \theta_{\tilde\nu}|_{\rm 1-loop} = 0.3$ as a function of 
          $m_{\tilde{\nu}_2}|_{\rm 1-loop}$.\label{fig:inverse}}
\end{figure}

Let $\hat{{\cal M}}^2_{\tilde{\nu}}(Q)$ be the running mass matrix
Eq.~(\ref{eq:sneutrino_tree}) at the renormalization point $Q$.  Then, the
pole mass matrix is given by
\begin{eqnarray}
  {\cal M}^2_{\tilde{\nu}} = \hat{{\cal M}}^2_{\tilde{\nu}}(Q) + \left( 
    \begin{array}{cc}
      \Delta M^2_{\tilde{\nu}LL}(Q) & \Delta M^2_{\tilde{\nu}LR}(Q) \\
      \Delta M^2_{\tilde{\nu}RL}(Q) & \Delta M^2_{\tilde{\nu}RR}(Q) \\
    \end{array}
  \right)\, ,
\end{eqnarray}
where $\Delta M^2_{\tilde{\nu}LL}(Q)$, $\Delta
M^2_{\tilde{\nu}LR}(Q)$, $\Delta M^2_{\tilde{\nu}RL}(Q)$ and $\Delta
M^2_{\tilde{\nu}RR}(Q)$ are radiative corrections to the LL, LR, RL
and RR components, respectively.  For large $\anu$ terms,
contributions to $\Delta M^2_{\tilde{\nu}}(Q)$'s are dominated by the
one--loop slepton--Higgs diagrams.
The main contribution to the light sneutrino mass shift results from
$\Delta M^2_{\tilde{\nu}RR}(Q)$, which can be aproximated as
\begin{eqnarray}
 \Delta M^2_{\tilde{\nu}RR}(Q)|_{\rm app}
 = \frac{\anu^2}{8 \pi^2} \left(\log \frac{m_{\tilde{\nu}_2}^2}{Q^2} - 1\right)\, .
\end{eqnarray}
Figure~\ref{fig:inverse} shows the the sneutrino soft terms $m_{\widetilde{L}}$,
$m_{\widetilde{N}}$ and $\anu$ that give a loop-corrected light sneutrino mass 
of $5$~GeV and mixing angle $\sin \theta_{\tilde\nu} = 0.3$
as a function of $m_{\tilde{\nu}_2}$.
The momentum and renormalization scales are set at $p = Q =
(m_{\tilde{\nu}_1} m_{\tilde{\nu}_2})^{1/2}|_{\rm 1-loop}$.
The figure also compares the complete 1-loop result (dashed green lines)
to the approximation
$m_{\tilde{\nu}_1}^2|_{\rm tree, app} =
m_{\tilde{\nu}_1}^2|_{\rm 1-loop} - \Delta M^2_{\tilde{\nu}RR}(Q)|_{\rm app}$ 
(dotted blue lines).

\subsubsection{Higgs mass corrections}

\begin{figure}[t!]\centering
   \includegraphics[width=0.6\textwidth]{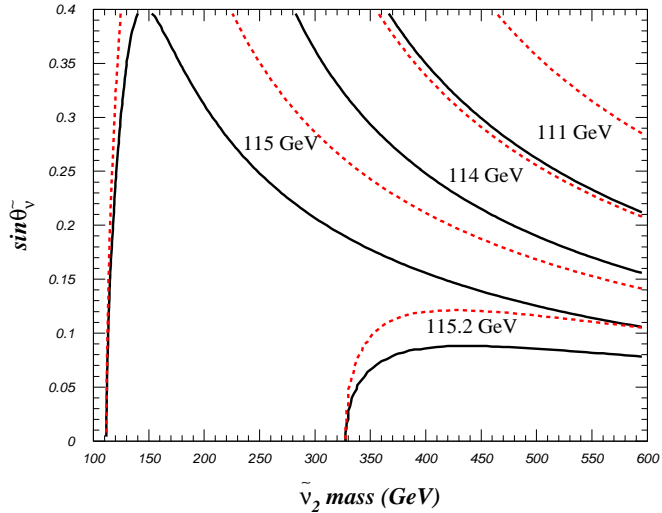}
   \vspace*{-2mm}
   \caption{ Contours of the radiatively-corrected light Higgs mass (from top to 
   bottom: $111$, $114$, $115$ and $115.2$~GeV) in the 
    $\sn$ versus $m_{\tilde\nu_2}$ plane, for $m_{\tilde\nu_1}=5$ GeV. 
    The dotted red lines are for the case of one light sneutrino, while the full black 
    lines are for three light sneutrino generations. \label{fig:mh}}
\end{figure}

After minimizing the Higgs potential, the mass of the lightest neutral Higgs reads
\begin{eqnarray}
  m_h^2 & = & m_Z^2 \sin^2 (\alpha + \beta) + m_A^2 \cos^2 (\alpha - \beta)
  \nonumber \\
  && +\, v^2 \left[ (\Delta \lambda_1 s_\alpha^2 c_\beta^2
  + \Delta \lambda_2 c_\alpha^2 s_\beta^2
  - (\Delta \lambda_3 + \Delta \lambda_4) 
  c_\alpha s_\alpha c_\beta s_\beta \right.
  \nonumber \\
  && \qquad \left. 
    + \Delta \lambda_5 (c_\alpha^2 c_\beta^2 + s_\alpha^2 s_\beta^2)
    - 2 (\Delta \lambda_6 s_\alpha c_\beta
    - \Delta \lambda_7 c_\alpha s_\beta) \cos (\alpha + \beta) \right] \, 
\label{eq:mh}
\end{eqnarray}
with $s_{\alpha}=\sin{\alpha}$, $c_{\alpha}=\cos{\alpha}$, etc.. 
The $\Delta\lambda_i$ include the radiative corrections to the quartic couplings. 
Loop diagrams involving sneutrinos can induce corrections to the quartic couplings 
through the presence of the weak scale $\anu$ term.   
If we neglect the Yukawa couplings of the sleptons as well as the trilinear terms for the charged
sleptons, then only $\lambda_2$ receives a correction of 
\begin{equation}
  \Delta \lambda_2^{(\widetilde{\nu})} = 
  - \frac{1}{16 \pi^2} \sum_{i=1}^{N_f}
  \frac{|A_\nu|^4}{(m_{\widetilde{\nu}_2}^2 - m_{\widetilde{\nu}_1}^2)^2}
  \left( 
    \frac{m_{\widetilde{\nu}_2}^2 + m^2_{\widetilde{\nu}_1}}{m_{\widetilde{\nu}_2}^2 - m_{\widetilde{\nu}_1}^2} 
    \log \frac{m_{\widetilde{\nu}_2}^2}{m_{\widetilde{\nu}_1}^2} -2 \right)\, ,
\end{equation}
with the sum running over the sneutrino flavours, $N_f=3$. 
Note that $\Delta \lambda_2^{(\tilde{\nu})}$ is negative, thus resulting in a decrease of the light 
Higgs mass Eq.~\ref{eq:mh}. We have checked that the effective potential technique gives the
same result as  PBMZ~\cite{Pierce:1996zz} when $p^2=0$. Setting $p^2=m_h^2$, we obtain the 
Higgs pole mass. We have incorporated the corrections to the light Higgs mass due to the sneutrinos 
in \suspect. 
When $\anu$ is large (which means large mixing and a large $m_{\tilde \nu_2}$) the radiative 
corrections from the sneutrino sector can drive the light Higgs mass below the LEP limit. 
This is illustrated in Fig.~\ref{fig:mh}, where we show  contours of constant $m_h$ in the $\sn$ 
versus $m_{\tilde\nu_2}$ plane. The dotted red lines are for the case of one light sneutrino, while the 
full black lines are for three degenerate light sneutrinos; in either case, $m_{\tilde\nu_1}=5$ GeV. 
Notice that the contour of $m_h=111$~GeV for three contributing sneutrino flavours almost falls 
together with the $m_h=114$~GeV contour of the one-sneutrino case.

\section{Constraints on the model}\label{sec:constraints}

\subsection{Collider constraints}

A light sneutrino with $m_{\tilde\nu}<m_Z/2$ will contribute to the invisible width of the Z boson, 
thus putting a constraint on the sneutrino mixing:  
\begin{equation}
    \Delta\Gamma_Z=\sum_{i=1}^{N_f} \Gamma_\nu\, \frac{\sin^4\theta_{\tilde\nu}}{2} 
      \left(1-\left(\frac{2 m_{\tilde\nu}}{m_Z} \right)^2 \right)^{3/2}< 2~{\rm MeV}
\end{equation}
where $\Gamma_\nu= 166$~MeV is the partial width into one neutrino flavour.
For one light sneutrino with $\mlsp=5$ (20) GeV, this leads only to a mild constraint on the  
mixing angle of $\sn<0.39$ ($0.43$). In the case of three degenerate sneutrinos, this constraint 
becomes stricter, $\sn<0.296$ ($0.33$).
 
We also impose the limits from SUSY \cite{lepsusy} and Higgs \cite{Schael:2006cr} searches at LEP2. 
Accounting for a theoretical uncertainty in the light Higgs mass of about 3~GeV, we require 
$m_h>111$~GeV. For a large value of the sneutrino mixing this implies an upper
bound on $\msnut$, see Fig.~\ref{fig:mh}. The radiative processes where a photon is emitted in addition to a pair of invisible
supersymmetric particles will contribute to the process
$e^+e^-\rightarrow \gamma+ invisible$, which has been searched for by the LEP2 experiments. 
Here invisible particles include not only the $\lsp$ but also
$\neuto$ or even $\neutt$ when they are the NLSP and NNLSP respectively as they decay in $\tilde\nu_1 \nu$.
As the  LEP2 limit we take $e^+e^- \rightarrow \gamma+invisible< 0.15$~pb at $\sqrt{s}=189-209$~GeV 
for $p_T^\gamma>0.02 \sqrt{s}$ and $\theta_{beam \gamma}>14$~deg~\cite{Achard:2003tx}.
We have computed the full 3-body cross-section for the single photon production using 
{\tt calcHEP}~\cite{Pukhov:2004ca} and found that it rarely exceeds tens of fb, which means 
it does not constrain our model.   
The reason for this is that for a large cross-section  it is necessary to
have a light particle exchanged in the t-channel. For neutralino production this means a light selectron, 
which is only possible in our 3 generation model.  Likewise, $\tilde\nu_e\tilde\nu_e\gamma$ production, 
which is enhanced by t-channel chargino exchange, is contributing only in the 3 generation case.

In \cite{Goodman:2010yf,Bai:2010hh} it was argued that the search for one-jet events with large 
missing transverse energy,  so-called monojets, could provide a stronger limit on light DM than 
current DD experiments. 
Monojet searches at the Tevatron look for events with leading jet $p_T>80$~GeV and missing
$E_T>80$~GeV, while 2nd jet $p_T<30$~GeV and more jets are vetoed. 
An analysis \cite{Aaltonen:2008hh} by the CDF collaboration of $1\,{\rm fb}^{-1}$ of data gave 
8449 events, with  an expected background of $8663\pm 332$. Using {\tt calcHEP}, we have 
computed the cross-section of $p\pbar\to \lsp\lsp+g$ at Tevatron energies and found that after cuts 
it is typically of the order of $0.1-1$~fb. Indeed for the parameter points that pass all other 
(including DD) constraints, see the scan of the following section, we find cross-sections of at 
most 1.5~fb. Thus the monojet search does not provide any additional constraint on the model.

In what follows, when we discuss DM allowed scenarios, it is implicitly understood 
that collider constraints are satisfied.

\subsection{Relic abundance of sneutrino}
 
For computing the sneutrino relic abundance, we assume the standard freeze-out picture. 
We do not consider non-thermal sneutrino production.  
This is justified because the mixed sneutrino has electroweak interactions. 
We have implemented the mixed sneutrino model in \micromegas, which allows for 
a fully automatic computation of the annihilation and DD processes.
Note that in the computation of the relic abundance we have not included the extra degrees 
of freedom corresponding to the RH neutrino. As these particles decouple early,  
this will only induce a correction at the few percent level on the effective degrees of freedom, 
which is negligible for our purpose.

The main annihilation channels for a light sneutrino are  
{\it i)}~$\lsp\lsp\to \nu\nu$ ($\lsp^*\lsp^*\to \bar\nu\bar\nu$)
through neutralino t-channel exchange,  
{\it ii)}~$\lsp\lsp^* \to b\bar{b}$ through exchange of a light Higgs in the s-channel, 
and {\it iii)}~$\lsp\lsp^* \to f\bar{f}$ through Z exchange.
The annihilation into neutrino pairs proceeds mainly through the wino component of the 
neutralino and is proportional to $\sin^4\theta_{\tilde\nu}$; it is largest for light winos. 
The Z exchange is also proportional to $\sin^4\theta_{\tilde\nu}$. 
The light Higgs exchange, on the other hand, is proportional to $(\anu\sn)^2$.  
Note in particular that for a fixed value of the sneutrino mixing angle, the Higgs contribution 
will increase with $m_{\tilde{\nu}_2}$ as $\anu$ also increases. 

\begin{figure}[t]\centering
\includegraphics[width=0.6\textwidth]{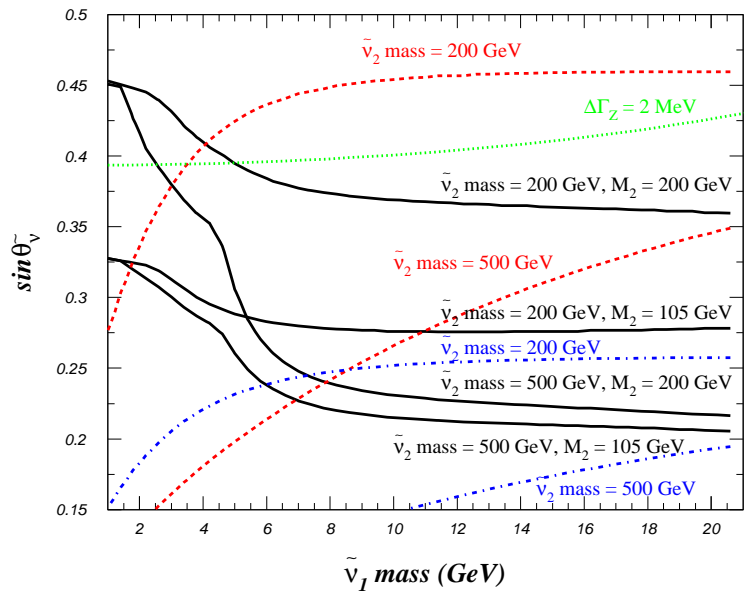}
\vspace*{-2mm}
\caption{In black (full lines), contours of $\Omega h^2=0.1$ in the $\sn$ versus $\mlsp$ plane 
for one light sneutrino, with  $\msnut=(200,\,500)$~GeV and  $M_2=(105,\,200)$~GeV. The 
remaining parameters are fixed as explained in Sec.~\ref{sec:spectrum}. The dashed red and 
dash-dotted blue lines show contours of constant $\sigma^{\rm SI}_{\lsp N}=10^{-4}$~pb and 
$10^{-5}$~pb, respectively. The dotted green line shows the limit from the Z width.}
\label{fig:omega}
\end{figure}

The behaviour of $\Omega h^2$ as a function of the sneutrino mass and mixing angle is 
displayed in Fig.~\ref{fig:omega} for the case of one light sneutrino. 
A larger mixing is required for light masses. This is related to the fact that $\Omega h^2$ 
is inversely proportionnal to the number of degrees of freedom ($g_{\rm eff}$). 
At the temperature where the QCD phase transition occurs, around 
$T_{\rm QCD}\approx 300{\rm MeV}$, the number of degrees of freedom starts to drop 
and $\Omega h^2$ increases. This is relevant for DM masses below ca.\ 6~GeV, 
where the freeze-out temperature $T_f\approx m_{\rm DM}/20$ is of the order of $T_{\rm QCD}$.  
Furthermore note that the uncertainty in the change of $g_{\rm eff}$ around $T_{\rm QCD}$ 
will induce some uncertainty in the computation of $\Omega h^2$. This is particularly important 
for $\mlsp < 2$~GeV because in this case $T_f \approx 100-150$~MeV, precisely where there 
is a sharp drop and a large uncertainty in $g_{\rm eff}$.
We have not considered these corrections to the relic abundance as only a few scenarios 
fall in this category. 

\begin{figure}[t]\centering
\includegraphics[width=7.4cm]{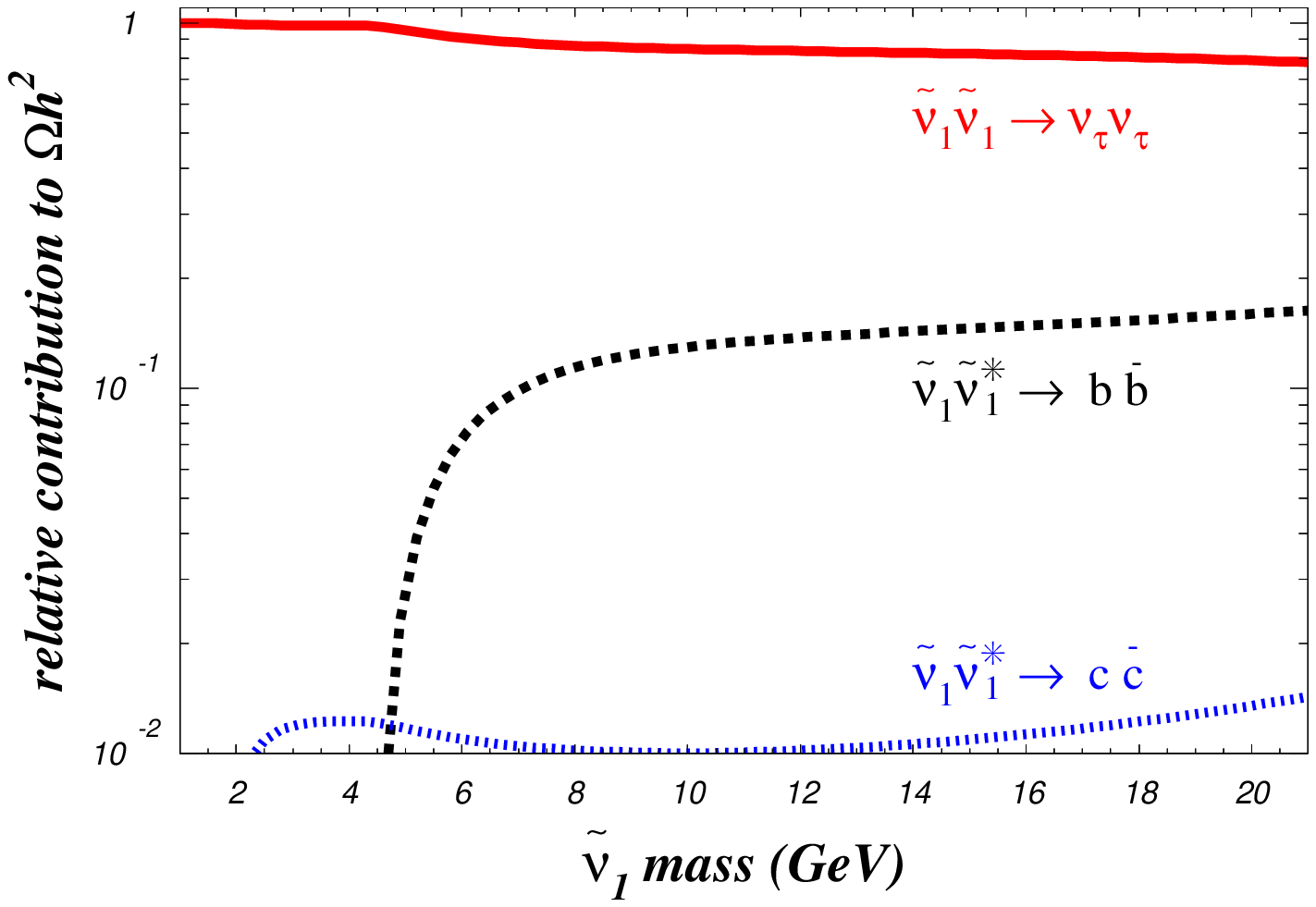}\quad
\includegraphics[width=7.4cm]{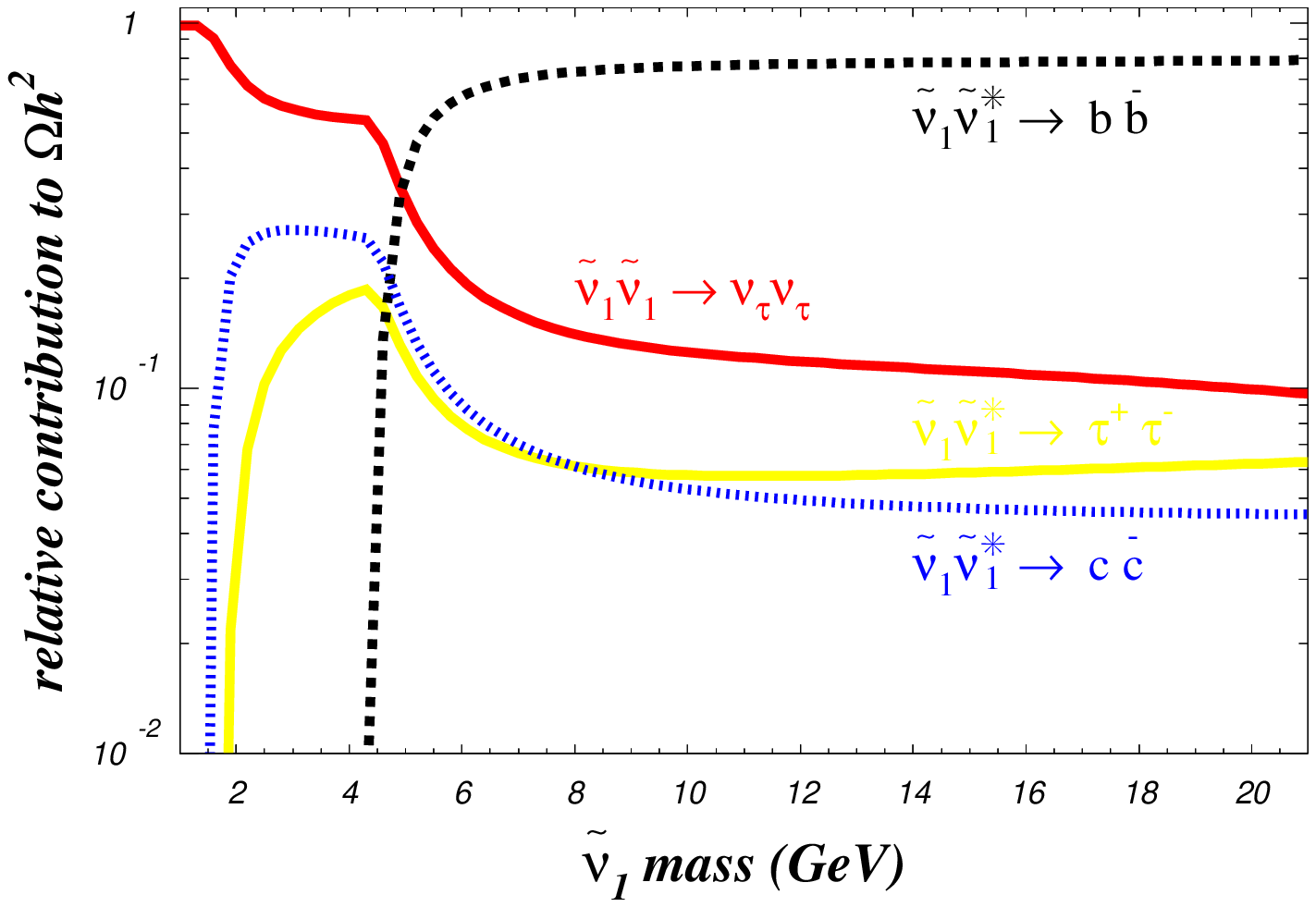}
\vspace*{-2mm}
\caption{Relative contributions of different annihilation channels as a function of the $\lsp$ mass, 
on the left for $\msnut=200$~GeV and $\sn=0.35$, on the right for $\msnut=500$~GeV and $\sn=0.22$, 
cf.\  Fig.~\ref{fig:omega}. In both plots, $M_2=200$~GeV.}
\label{fig:omegaContrib}
\end{figure}

The dependence of  $\Omega h^2$ on the gaugino mass, $M_2$ is also displayed in Fig.~\ref{fig:omega}.
A lower mass requires a smaller mixing, this is because the self-annihilation channel into neutrinos 
is increased in this case.  
The relative contributions of the various annihilation channels are shown in 
Fig.~\ref{fig:omegaContrib} for the two cases of {\it a)} large mixing, $\sn=0.35$, but relatively 
small $\anu$ (left panel) and {\it b)} smaller mixing, $\sn=0.22$, but large  $\anu$ (right panel).
In the large $\anu$ case,  the contribution from the Higgs exchange enhances the $b\bar{b}$ channel if kinematically accesssible.  

\subsection{Direct detection}

The spin-independent (SI) scattering of $\lsp$ on nucleons occurs through Z or Higgs exchange. 
The Z exchange is again suppressed by the sneutrino mixing angle. The scattering cross-section 
on a nucleus due to Z exchange is given by 
\begin{equation}
\sigma^{\rm SI,\,Z}_{\lsp N} =\frac{G_F^2}{2\pi} \mu_\chi^2  \left((A-Z)-(1-4\sin^2\theta_W)Z\right)^2 \sin^4\tsn \,,
\label{eq:dd}
\end{equation}
where $\mu_\chi$ is the sneutrino--nucleus reduced mass, while $A$ is the atomic weight and $Z$ 
the number of neutrons of the nucleus. 
One peculiarity of the Z-exchange contribution is that the proton cross-section is much smaller than 
the neutron one, with the ratio of amplitudes $f_p/f_n=(1-4\sin^2\theta_W)$. 
The Higgs contribution on the other hand, which becomes dominant for large values of $\anu$, 
is roughly the same for protons and neutrons, 
\begin{equation}
\sigma^{{\rm SI,}\,h}_{\lsp N} =\frac{\mu_\chi^2 }{4\pi}  \frac{g_{{h\lsp\lsp}}^2 } {m_h^4 m_{\lsp}^2} 
\left( (A-Z) \sum_q g_{hqq} f_q^n m_n + Z \sum_q g_{hqq} f_q^p m_p   \right)^2  \,,
\label{eq:ddh}
\end{equation}
where $g_{hqq}=e/(2 M_W s_W) x_q $  ($x_u= -\cos\alpha/\sin\beta, x_d=\sin\alpha/\cos\beta$)
 is the Higgs coupling to quarks after the quark  mass has been factored out, and
$g_{h\lsp\lsp}$ is the coupling to the LSP as given in Appendix A.

The total SI cross-section is obtained after averaging over the $\lsp N$ and ${\lsp}^*N$ cross 
sections, where we assume equal numbers of sneutrinos and anti-sneutrinos. Here note that 
the interference between the Z and Higgs exchange diagrams has opposite sign for 
$\lsp N$ and ${\lsp}^*N$, leading to an asymmetry in sneutrinos and anti-sneutrinos scattering 
if both Z and Higgs exchange are important. We will come back to this in Section~\ref{sec:neutrinos}, 
when we discuss signals from sneutrinos captured in the Sun. 

A comment is in order concerning theoretical uncertainties. 
The computation of the elastic scattering cross-section for the Higgs-exchange diagram 
depends on the quark coefficient in the nucleons, which 
can be determined from  the pion-nucleon sigma term,   $\sigma_{\pi N}$ and from the SU(3) 
symmetry breaking effect, $\sigma_0=35\pm 5$~MeV~\cite{Gasser:1990ce}. 
By default we take  $\sigma_0=35$~MeV and $\sigma_{\pi N}=45$~MeV~\cite{Gasser:1990ce}. 
This leads to
\begin{eqnarray}
\label{eq:scalar}
f^p_{d}=0.026\,, \;\; f^p_{u}=0.020\,, \;\; f^p_{s}=0.13\,, \nonumber\\
f^n_{d}=0.036\,, \;\; f^n_{u}=0.014\,, \;\; f^n_{s}=0.13\,.
\end{eqnarray}
More recent estimates of the pion-nucleon sigma term typically indicate 
larger values of $\sigma_{\pi N}=55-73$~MeV~\cite{Pavan:2001wz}.
The very recent lattice results also tend towards a larger value for the strange quark content of 
the nucleon, although uncertainties are still large~\cite{Ohki:2008ge,Babich:2009rq}.
The overall theoretical uncertainty that arises from the uncertainty in the scalar coefficients is 
relevant only for cases where the Higgs-exchange contribution dominates, since for the Z 
contribution the vector coefficients are simply determined by the valence quark content in the 
nucleon. Since the values in Eq.~(\ref{eq:scalar}) are rather on the low side, the DD cross-section 
shows a larger upward than downward variation when $\sigma_{\pi N}$ and $\sigma_0$ are varied 
within their allowed ranges. For example, for $(\sigma_{\pi N},\,\sigma_0)=(70,\,30)$~MeV
the cross-section can increase by up to a factor $3.5$ when Higgs exchange dominates, while
for $(55,\,35)$~MeV the increase is at most a factor $1.7$. The choice $(45,\,40)$~MeV 
on the other hand implies a decrease in the cross-section that can reach 30\%.

The limits on $\sigma_p^{\rm SI}$ from DD experiments are extracted from the observed 
limit on the LSP--nucleus scattering cross-section assuming that amplitudes for protons 
($f_p$)  and neutrons ($f_n$) are equal. In our model this is not the case when Z exchange 
dominates. Therefore, we compute instead the normalized cross-section on a point-like nucleus,  
\begin{equation}
\sigma^{\rm SI}_{\lsp N}= \frac{4 \mu_\chi^2}{\pi}\frac{\left( Z f_p+ (A-Z)f_n\right)^2}{A^2 }
\end{equation}
where the average over $\lsp$ and $\lsp^*$ is assumed implicitly. 
This cross-section can be directly compared with the limits on $\sigma^{\rm SI}_p$ 
given by the experiments. For Xenon $A=131, Z=54$ while for Germanium $A=76, Z=32$.

The SI scattering cross-section mainly depends on the three parameters of the 
sneutrino sector as long as $m_A$ is large enough so that heavy Higgs exchange can 
be neglected. Contours of constant $\sigma^{\rm SI}_{\lsp N}$ in the  $\sn$ versus $\mlsp$ 
plane are displayed in Fig.~\ref{fig:omega} for two different values of $m_{\tilde\nu_2}$.
For $\msnut = 200$~GeV, in the region $\mlsp \gsim 6$~GeV the $\anu$ parameter is
so small that scattering proceeds mainly via the Z-boson exchange; the
detection rate is almost independent of $\mlsp$. 
For  $\msnut = 500$~GeV, $\anu$ is so large that
the nucleon scattering has an important component from the Higgs-boson exchange; for this channel
the detection rate is proportional to $\mlsp^{-2}$. Therefore the detection rate
gets smaller for larger $\lsp$ mass.  This is a characteristic of 
scalar dark matter particles. Similarly  for $\mlsp < 6$~GeV, the nucleon scattering 
proceeds mainly through Higgs-boson exchange implying an increase for lower $\lsp$ mass.  \\

Before we proceed, another comment is in order concerning the $\tan\beta$ dependence. 
So far we have only considered $\tan\beta=10$. The main effect of increasing $\tan\beta$ 
is a lower $\stau_1$ mass for the same choice of $\msnu,\ \msnut,\ \sn$ and $\mu$. 
In our approach, this can be translated into an upper limit on $\tan\beta$ as a function 
of $\msnut$, or vice-versa a lower limit on $\msnut$ as a function of $\tan\beta$. 
In fact, for $\msnut=200$~GeV like in 
Figs.~\ref{fig:omega} and \ref{fig:omegaContrib}, one can only go up to $\tan\beta\approx 20$; 
for higher values one first violates the $\stau_1$ mass bound and then gets tachyonic staus. 
For high values of, e.g., $\tan\beta=50$, one needs $\msnut\gsim 300$~GeV. 
Another effect is a slightly higher $h^0$ mass for higher $\tan\beta$. 

The influence on the relic density of the $\lsp$ is small, the main effect being 
a slight suppression of the $\lsp\lsp^*\to\tau^+\tau^-$ channel through the interference 
of the Higgs and chargino exchange diagrams. To give a concrete example, 
for $\mlsp=10$~GeV, $\msnut=500$~GeV, $\sn=0.22$, $M_2=200$~GeV and $\tan\beta=10$ 
we have $\Omega h^2=0.12$ with 6\% contribution from $\lsp\lsp^*\to\tau^+\tau^-$, cf.~Fig~\ref{fig:omegaContrib}.
For the same parameters but $\tan\beta=50$, we get $\Omega h^2=0.131$ with 4\% contribution 
from $\lsp\lsp^*\to\tau^+\tau^-$. Lowering $M_2$ to $105$~GeV leads to $\Omega h^2=0.092$ 
($0.1$) for $\tan\beta=10\ (50)$.

The SI scattering cross-section is affected by $\tan\beta$ through the Higgs mass 
and the bottom Yukawa coupling, but this effect is also small. In the above example 
with $M_2=200$~GeV, $\si_{\lsp p}=8.63\times 10^{-5}$~pb and $\si_{\lsp n}=1.17\times 10^{-4}$~pb 
change to $\si_{\lsp p}=8.04\times 10^{-5}$~pb and $\si_{\lsp n}=1.11\times 10^{-4}$~pb 
when increasing $\tan\beta$ from 10 to 50. Here note that $m_h=114.2\ (115.2)$~GeV 
and $\anu=310.7\ (309.2)$~GeV at $\tan\beta=10\ (50)$.

We conclude that the choice of $\tan\beta$ has little influence on the results. 
However, lower $\tan\beta$ has more parameter space in the sense that one can go to 
lower $\snt$ masses. Therefore in the following we will focus on $\tan\beta=10$.

\section{Results for one light sneutrino flavour}\label{sec:onesnu}

For the light sneutrino to be a viable dark matter candidate, in addition to the collider constraints, 
we require that both the $3\sigma$ upper bound from WMAP5 \cite{Dunkley:2008ie}, 
$\Omega h^2< 0.1285$, and the DD limits be satisfied. In fact, DD here provides the more 
stringent constraint.  
We first discuss the case of one light sneutrino, concentrating on the third generation. 

For the numerical analysis, we perform a random scan of the parameters of the 
sneutrino and gaugino sector. To search efficiently the region with a light sneutrino, we use 
$m_{\tilde\nu_{1,2}}$ and $\sn$ as input, from which we compute $m_{\widetilde{L_3}}$, 
$m_{\widetilde{N_3}}$ and $\anu=A_{\tilde\nu_3}$.  
The parameter ranges we scan over are 
\begin{equation}
  \begin{array}{lll}
     1~{\rm GeV} < m_{\lsp} < 15~{\rm GeV}\,, & \, &  100~{\rm GeV} < m_{\tilde\nu_2} < 1000~{\rm GeV}\,, \\
     0< \sn <0.5 \,, & & 100~{\rm GeV} < M_2=2M_1 < 500~{\rm GeV} \,.
  \end{array}
\end{equation}
All other parameters have little influence on the light sneutrino scenario 
and are therefore fixed as specified in Section~\ref{sec:spectrum}. In particular, the stau masses are 
determined through $m_{\widetilde R_3}=m_{\widetilde L_3}$ and $A_\tau=0$, while all other soft 
masses are set to 1~TeV. Moreover, having checked that it does not influence the results, we 
set $\mu=800$~GeV.  
The scan is carried out in two steps with 50000 points each, the first varying $M_2$ from 
100 to 150 GeV (light wino case) and the second varying $M_2$ from 150 to 500 GeV (heavy wino case).  
Extending the upper value for the range of $M_2$ has no influence on our results. 

\begin{figure}[t]\centering
\includegraphics[height=6.6cm]{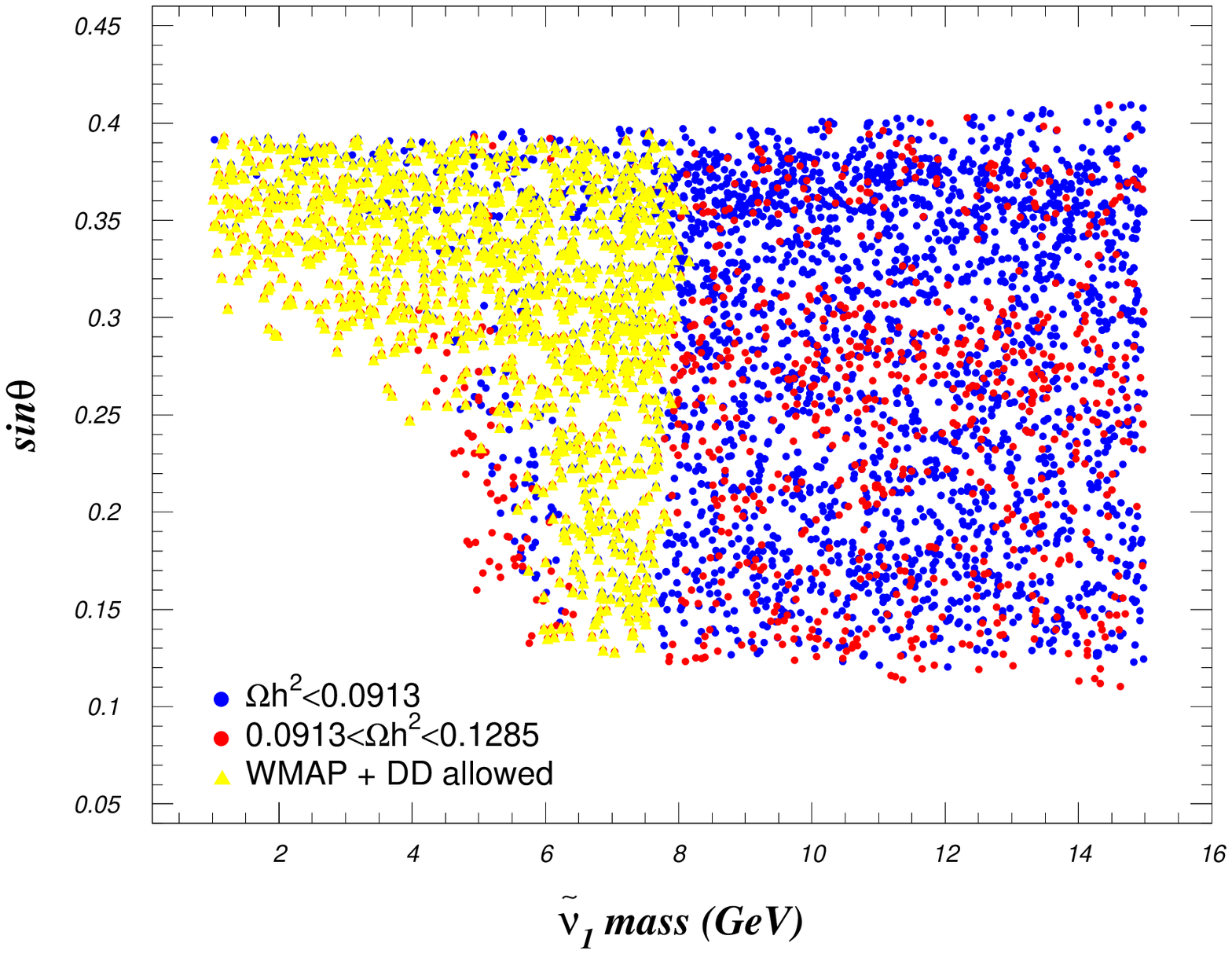}\hspace*{-1mm}
\includegraphics[height=6.4cm]{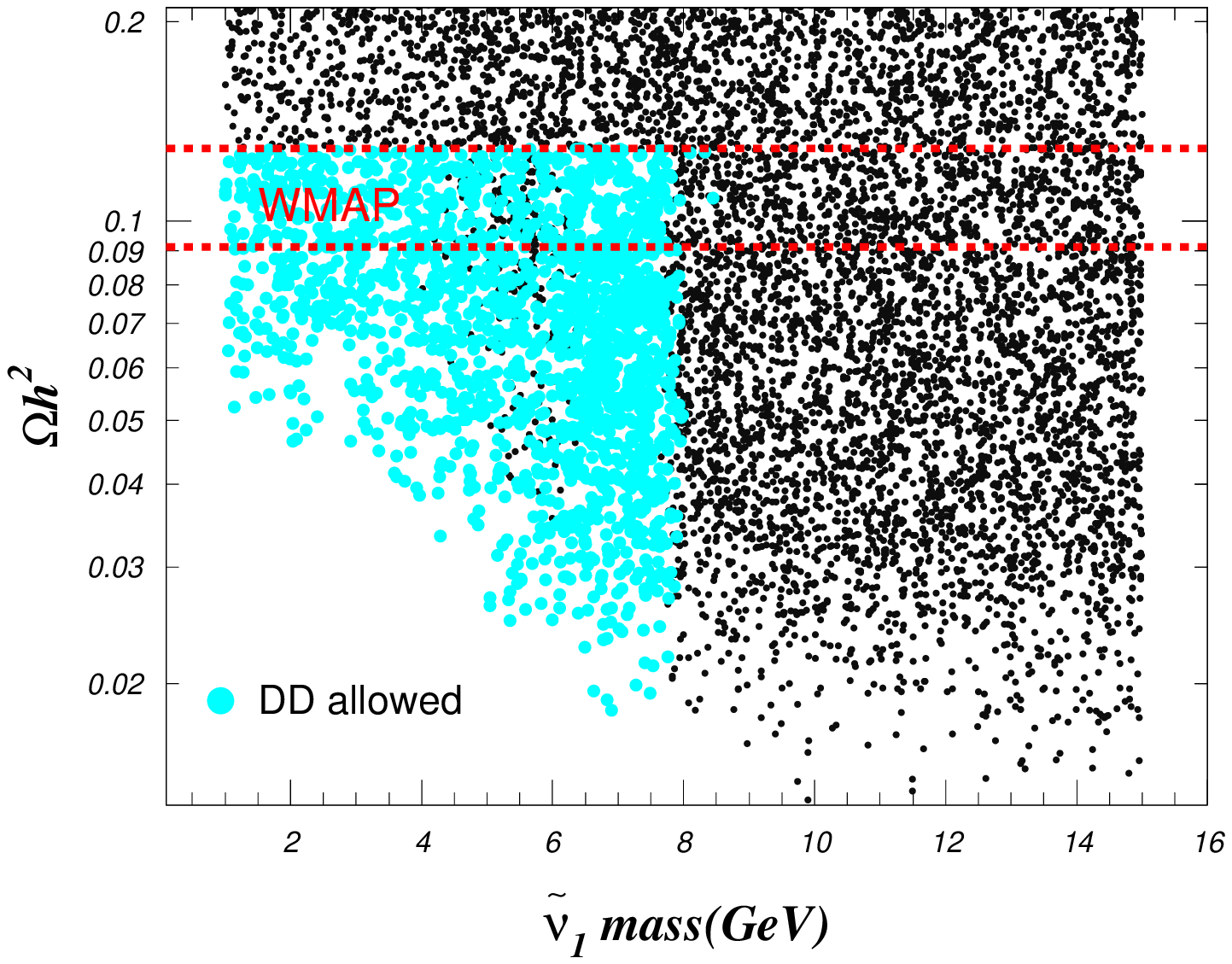}
\caption{Scan results for the one light sneutrino case. 
On the left, a scatter plot of $\sn$ versus $\mlsp$. Here the red (blue) dots show models that have 
a relic density within (below) the $3\sigma$ WMAP range. The yellow triangles refer to models that 
satisfy the WMAP upper bound as well as the DD limits from CoGeNT and Xenon10. 
The plot on the right shows $\Omega h^2$ versus $\mlsp$ with the light blue points passing the  
DD constraints.}
\label{fig:good1} 
\end{figure}

The scan results are shown in Fig.~\ref{fig:good1} in the 
$\sn$ versus $\mlsp$ and $\Omega h^2$ versus $\mlsp$ planes.  
We find scenarios that satisfy the WMAP and DD limits for sneutrinos as light as 1~GeV. 
Such light $\lsp$'s below ca.\ 4~GeV require a large mixing, $\sn\gsim 0.25$,  and annihilate 
predominantly into neutrinos. A more modest mixing of $\sn> 0.12$ is needed for $\lsp$ 
masses above the $b$-threshold, where annihilation into $b\bar{b}$ through Z or $h$ can 
contribute significantly. 

\begin{figure}[t]\centering
\includegraphics[width=10cm]{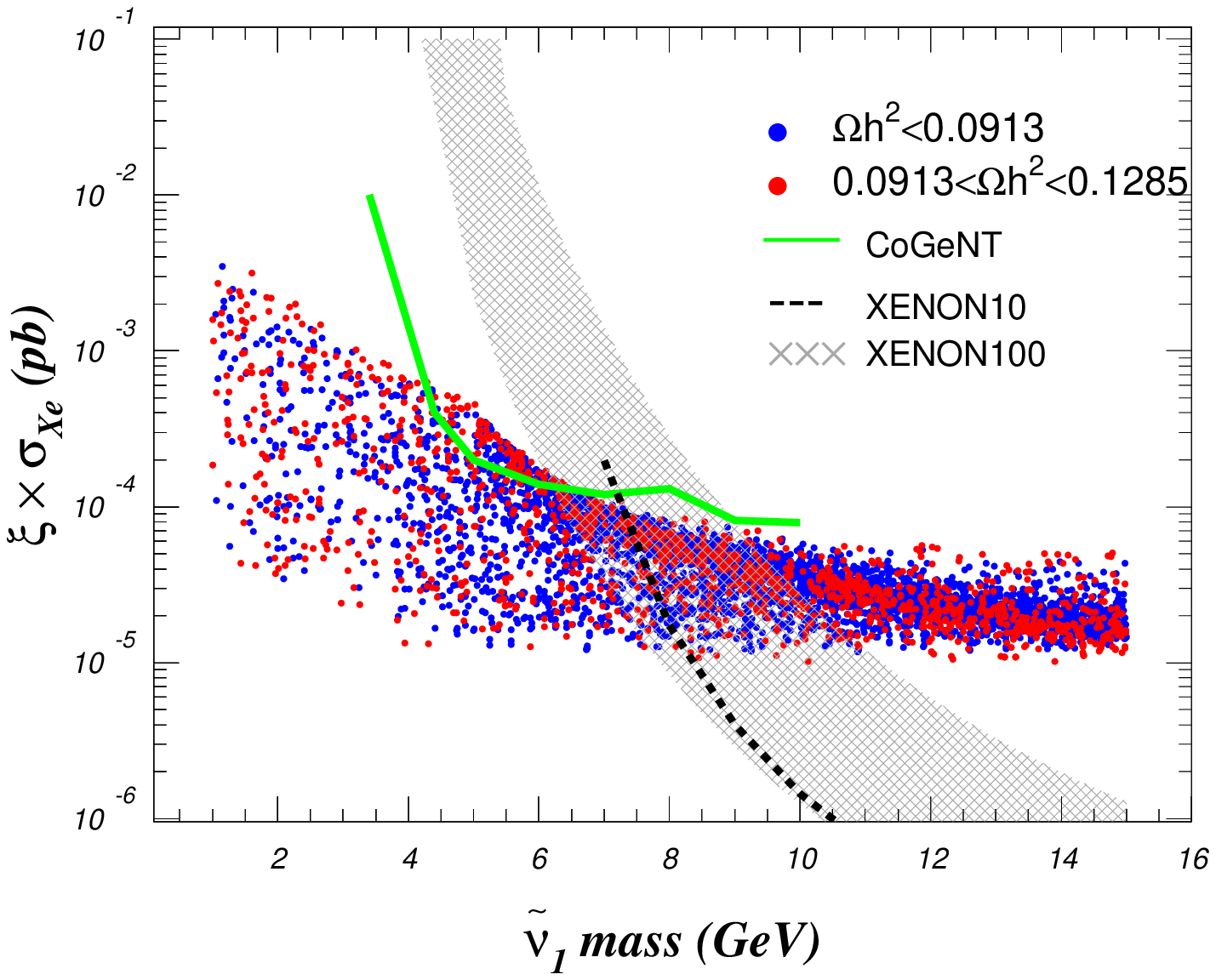}
\vspace*{-2mm}
\caption{Scatter plot of re-scaled $\sigma^{\rm SI}_{\lsp Xe}$ as a function of $\mlsp$ for the case 
of one light sneutrino. The red (blue) dots show models that are within (below) the WMAP range. 
Also shown are the limits from CoGeNT (green), Xenon10 (dashed), and Xenon100 limit (crosses).  
The CoGeNT and Xenon10 limits are for $v_0=220$~km/s, $v_{esc}=600$~km/s and 
$\rho=0.3$~GeV/cm$^3$. 
The Xenon100 band is obtained by varying $v_0=180$--$260$~km/s, $v_{esc}=500$--$600$~km/s, 
$\rho=0.2$--$0.6$~GeV/cm$^3$ and an $L_{\rm eff}$ that decreases at low energies, 
see Sec.~\ref{sec:uncertainties} for details. 
\label{fig:good1dd}}
\end{figure}

Figure~\ref{fig:good1dd} shows the expected SI cross-sections for $Xe$ as a function of the $\lsp$ mass, 
together with the limits from CoGeNT, Xenon10 and Xenon100, which give the most stringent 
constraints on light DM~\cite{Aalseth:2010vx,Angle:2007uj,Aprile:2010um}. The cross-section is appropriately 
re-scaled if the $\lsp$ is only a part of the DM, i.e. $\xi=\Omega h^2/0.11$ for points with 
$\Omega h^2<0.0913$ and $\xi=1$ otherwise. 

When the sneutrino annihilation is dominated by Z or Higgs exchange, the prediction for the 
SI scattering cross-section, which is also dominated by the Z or Higgs exchange diagrams, 
is directly related to the annihilation cross-section. In this case, after re-scaling for the lower 
dark matter density if $\Omega h^2<0.0913$, $\ssi$ varies only within a factor of 2  
(not taking into account uncertainties due to the quark coefficients Eq.~(\ref{eq:scalar})). 
On the other hand, when the sneutrino annihilates dominantly into neutrino pairs, there is no such 
correlation between $\Omega h^2$ and $\ssi$, and it is possible to suppress the direct detection 
cross-section by more than one order of magnitude. These scenarios correspond to the scatter 
points with low cross-sections in Fig.~\ref{fig:good1dd}. 

The points that successfully pass all constraints have sneutrino masses  
of about 1 to 8 GeV and include many scenarios that have cross-sections within the range 
favoured by CoGeNT. In particular the bulk of our scenarios with masses around 7--8 GeV 
are in the region allowed by CoGeNT and Xenon. 
Recall also that such light DM with a cross-section of a few times $10^{-5}$~pb could 
cause the events observed by CRESST. 
For $\mlsp\approx 5$~GeV, many of the WMAP-allowed scenarios lie above the CoGeNT limit. 
A value of $M_2\sim 100-120$~GeV is required in this case to achieve efficient enough 
annihilation into neutrinos. For lower masses, the DD limit becomes much weaker, leaving  
the relic density and Z width as the main constraints. 
Moreover, we recall that, while some fine-tuning is needed to achieve a light 
$\lsp$, there are no large hierarchies among the soft terms (cf.~Section~\ref{sec:model1}). 
In particular the ratio between $\anu$ and $M_2$ roughly ranges from $1/5$ to 5~(8) for 
$\mlsp\lsim 6\ (8)$~GeV.

It is also intriguing that in Fig.~\ref{fig:good1dd} there is a lower limit on the DD cross-section 
of about $10^{-5}$~pb, which is almost flat for $\mlsp\gsim 4$~GeV. 
It arises because the lowest $\ssi$ is obtained when the dominant contribution 
to the relic abundance is annihilation into neutrinos. In this case, scattering off nucleons 
proceeds dominantly through the Z-boson exchange, which is almost independent of 
$\mlsp$. Note that this lower limit is free of the theoretical uncertainties on the quark coefficients in
the nucleon.
The mild mass dependence of the lower limit on $\ssi$ for $\mlsp < 4$~GeV arises 
because the Higgs exchange which is dominant in this region,  is porportional to 
$1/\mlsp^2$, Eq.~(\ref{eq:dd}). 
Overall, this lower limit means that mixed sneutrino DM with a mass of a few  
GeV will be either discovered or excluded if the DD experiments can cover SI cross-sections 
down to $10^{-5}$~pb. 
 
\begin{figure}[t]\centering
\includegraphics[width=10cm]{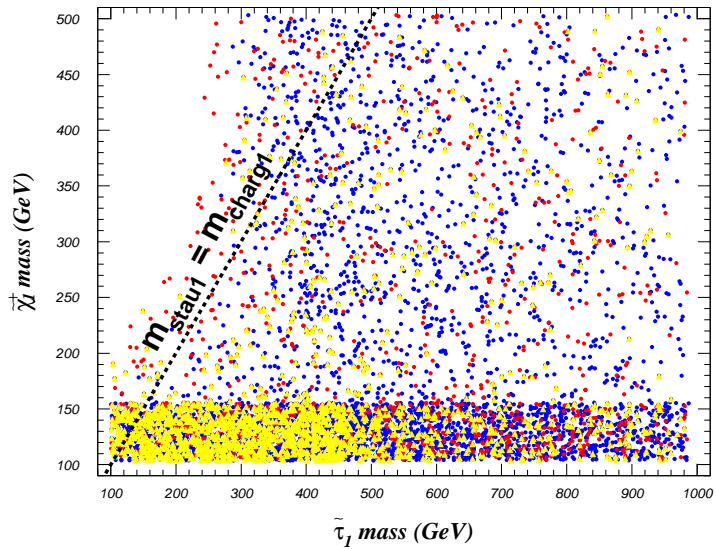}
\vspace*{-2mm}
\caption{Same as the lhs plot of Fig.~\ref{fig:good1} but in the $\mchar$ versus $m_{\tilde\tau_{1}}$ plane.
The red (blue) dots denote points that have a relic density within (below) the $3\sigma$ WMAP range
but too large a DD cross-section. The yellow triangles denote points that satisfy the WMAP upper 
bound as well as the DD limits from CoGeNT and Xenon10. \label{fig:good1staus}}
\label{fig:stau}
\end{figure}

The mass of the ${\tilde\nu_2}$ and consequently of the $\tilde\tau_{1,2}$ is also constrained. 
When $M_2$ is small, around ca.\ $100$~GeV, and annihilation into neutrinos dominates,  
the angle $\sn>0.25$ to provide sufficient annihilation but the mass of ${\tilde\nu_2}$ is not constrained. 
On the other hand when Higgs exchange is dominant, the annihilation is proportionnal to 
$\anu^2 \sn^2 \approx\msnut^4 \sn^4$, which gives  a lower bound on $\msnut$. 
Consequently a lower bound on  $m_{\tilde\tau_{1}}\gsim 250$~GeV is obtained for large $M_2$.  
This is illustrated in Fig.~\ref{fig:good1staus}, which shows the points of Fig.~\ref{fig:good1} 
in the $\mchar$ versus $m_{\tilde\tau_{1}}$ plane. We found a few scenarios 
with  $m_{\tilde\tau_{1}}<\mchar$. 
For none of these points $\neutt$ or $\charg$ have a significant branching fraction into 
slepton/lepton pairs. This has important consequences for collider searches, as we will 
discuss in Section~\ref{sec:colliders}. 

\begin{figure}[t]\centering
\includegraphics[width=10cm]{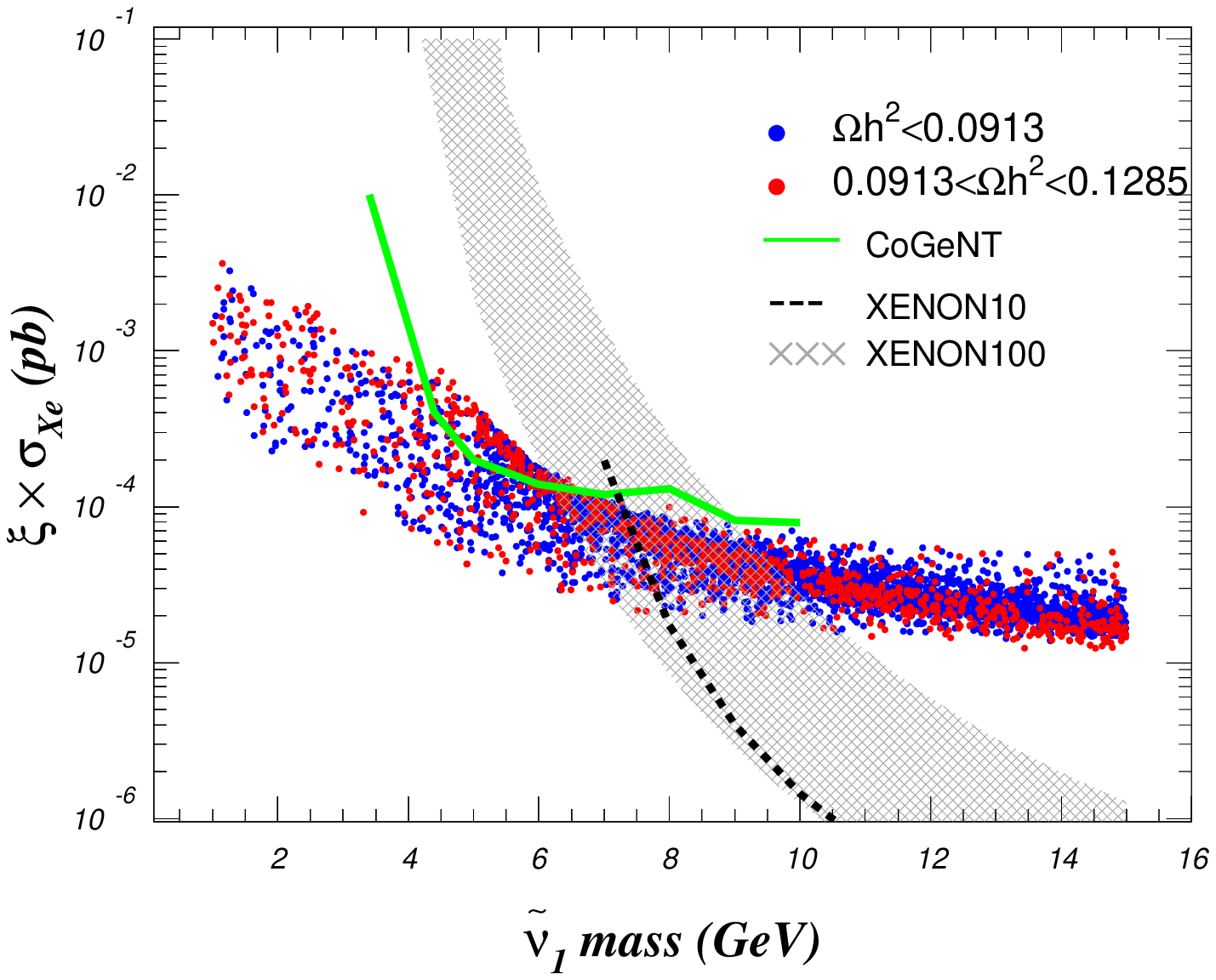}
\vspace*{-2mm}
\caption{Same as Fig.~\ref{fig:good1} but for $\tan\beta=50$.
\label{fig:good1dd50}}
\end{figure}

Let us briefly come back to the $\tan\beta$ dependence. 
We have argued above that $\tan\beta=10$ has a larger parameter space than 
higher values, but otherwise results are very similar. This is illustrated in 
Fig.~\ref{fig:good1dd50}, which shows scan results as in Fig.~\ref{fig:good1} 
but for $\tan\beta=50$. As can be seen, the only change is that points with 
low DD cross sections, corresponding to a low $\msnut$, are removed. 
Indeed the scan points for $\tan\beta=50$ that pass all constraints are 
just a subset of those for $\tan\beta=10$. In the following we will therefore 
use the $\tan\beta=10$ scan points. 

\subsection{Uncertainties in the DD limits}\label{sec:uncertainties}

As the DD limit plays a crucial role in constraining the parameter space, it is important to 
consider the uncertainties involved in extracting the limit on $\sigma^{SI}_{\chi p}$ from experiments.
One source of uncertainty is the assumed DM velocity distribution. 
In particular, there is a minimum velocity necessary to pass the energy threshold of the detector,  
$v^2_{min}=\frac{m_N E_R}{2\mu_\chi^2}$. Thus the direct detection limit for low masses is 
particularly sensitive to the upper part of the velocity distribution.  To illustrate this effect, 
we assume an isothermal velocity distribution and allow for a 2$\sigma$ variation of $v_0=220\pm 20\rm{km/s}$.
We also vary  the  velocity above which DM will escape from the Galaxy in the range 
$500<v_{esc}<600\rm{km/s}$~\cite{Smith:2006ym}.\footnote{Several velocity distributions were 
considered in~\cite{Bottino:2005qj,Bottino:2007qg}. Varying the input parameters
of the isothermal distribution as we do, reproduces most of the span of variation of the limits 
found in those studies.}
The local dark matter density is another factor that possesses a large uncertainty, it is generally 
assumed to be $\rho=0.3$~GeV/cm$^3$ but can vary from $0.2-0.7$. 
(For a review of astrophysical uncertainties on the velocity and density distribution see~\cite{Green:2010ri}.) 

Finally, there is an additional uncertainty that is specific to Xenon, it comes from the scintillation factor 
$L_{\rm eff}$ that allows to convert the measured electron energy into the nuclear recoil energy. 
In particular the threshold for nuclear recoil energy strongly depends on the scintillation factor
at low energies. New measurements have shown that the scintillation efficiency can be  significantly 
lower than the value used in Ref.~\cite{Angle:2007uj}---and this especially at low energies---thus 
weakening the limit on $\sigma^{SI}_{\chi p}$~\cite{Angle:2009xb}. The Xenon10 limit in Fig.~\ref{fig:good1dd} 
uses a constant value for $L_{\rm eff}$ while the band for Xenon100~\cite{Aprile:2010um} uses a value 
of  $L_{\rm eff}$ that decreases at low energies~\cite{Manzur:2009hp}.

\section{Results for three degenerate light sneutrinos}\label{sec:threesnu}

In the case of three light sneutrinos, the constraint from the Z width on the
sneutrino mixing angle is stronger. Moreover, for three exactly degenerate sneutrinos, 
the relic density can increase by a factor up to 3 as compared to the one-generation case.
This is because the effective annihilation cross-section including the coannihilation channels is 
$\langle \sigma v\rangle \propto (\sum g_i^2 \sigma_{ij})/\sum g_i^2$  where $g_i$ is the number of 
degrees of freedom and $\sigma_{ij}$ the cross-section for annihilation of two supersymmetric 
particles, $\chi_i,\chi_j$ into SM particles.
Thus when annihilation is dominated by the processes $\tilde\nu_i \tilde\nu_i^*\to f\bar f$,  
$\langle \sigma v\rangle$ increases by a factor of 3 as compared with the one generation case. 
This then leads to  a tension between the relic density constraint which requires significant mixing 
and the Z width which strongly constrain this mixing angle. 
The relic density constraint is particularly severe at low masses where we had 
$\Omega h^2\approx 0.11$ in the one generation case. The smaller mixing can of course 
be partially compensated by $\anu$ however this then typically violates the DD limits. 
Furthermore, for three light sneutrinos with large $\anu$, the Higgs mass constraint becomes 
much more severe.

Decreasing $M_2$ makes it easier to satisfy the relic density constraint. Indeed in this case 
annihilation is dominated by production of neutrinos and all channels 
$\tilde\nu_i \tilde\nu_j\rightarrow \nu_i\nu_j$ contribute equally to the annihilation 
cross-section so that $\Omega h^2$ is only a factor 3/2 larger than 
for the one generation case.

\begin{figure}[t!]\centering
\includegraphics[width=10cm]{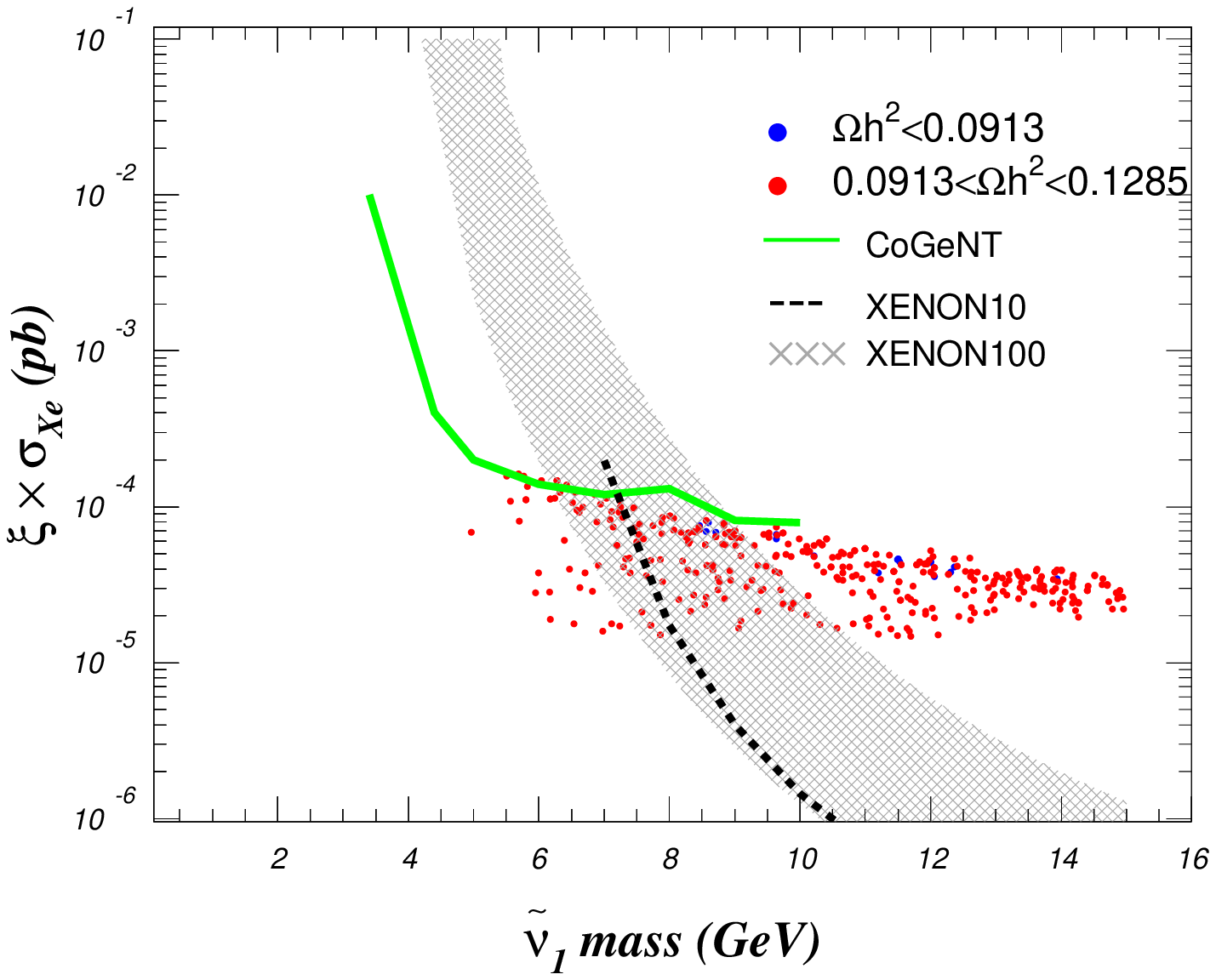}
\caption{Scatter plot of $\sigma^{\rm SI}_{\lsp Xe}$ versus $\mlsp$ analogous to 
Fig.~\ref{fig:good1dd} but for three degenerate light sneutrinos. 
\label{fig:good3dd}}
\end{figure}

We perform a random scan analogous to the one in the previous section. 
The result of this scan is shown in Fig.~\ref{fig:good3dd}. 
As expected, the parameter space for sneutrino DM is now much more restricted, 
and only few points pass all constraints. 
The allowed sneutrino masses now range from 4.5 to 8~GeV. 
For most scenarios the  chargino mass lies just above the LEP limit.
The sneutrino mixing is around $\sn \approx 0.27-0.29$, while the slepton masses are 
restricted to  $m_{\tilde{\ell}_{L,R}}\approx m_{\tilde{\nu}_2} \approx 150-480$~GeV.
This is because larger masses induce too large corrections to $m_h$, 
while smaller masses reduce the value of $\anu$ and thus the Higgs-exchange contribution 
to the annihilation processes. 
It is also interesting to note that in the DD-allowed region there are no points with $\Omega h^2$ 
below the $3\sigma$ WMAP range (and overall there are only very few points with 
$\Omega h^2<0.0913$).

The exact mass splitting between sneutrinos of different flavours strongly influences the
allowed parameter space. For example, a 1 GeV mass splitting is enough to suppress 
any coannihilation contributions to a negligible level. Such a mass splitting can be induced by 
small splittings in the soft terms, which are rather generic in supersymmetric models even if one 
starts out with universal soft terms at a high scale. Taking, for instance, 
$m_{\widetilde L_1}=m_{\widetilde L_2}=m_{\widetilde L_3}$, 
$m_{\widetilde N_1}=m_{\widetilde N_2}=m_{\widetilde N_3}$ and just 1\% difference 
in the $\anu$ terms, $A_{\tilde\nu_{e,\mu}}=0.99A_{\tilde\nu_\tau}$, we find mass splittings of 2--14~GeV.
In this case, the heavier $\tilde\nu_{1e,\mu}$ decay into the $\tilde\nu_{1\tau}$ LSP 
through 3-body modes~\cite{Kraml:2007sx} with decay widths ranging from $10^{-12}$ to 
$10^{-8}$~GeV. The dominant mode is into neutrinos; visible decays, e.g.,  
$\tilde\nu_{1e}\to e^\mp\tau^\pm\tilde\nu_{1\tau}$, have at most few percent branching ratio. 
In this case we are back to the case of one light sneutrino presented  previously with only a more 
severe constraint on $\sn$ and $\anu$ from the Z width and light Higgs mass. 
In what follows we will therefore concentrate on the case of one light sneutrino.

\section{Indirect detection}\label{sec:indirect}

\subsection{Neutrinos from annihilation in the Sun}\label{sec:neutrinos}

Neutrinos originating from annihilation of DM captured in the Sun could provide a good signature 
for the mixed sneutrino model. In scenarios with light sneutrinos, a detector with a low threshold 
is needed. The Super-K detector has a threshold  of $1.6$~GeV, while large detectors like Antares 
or ICECUBE have threshold energies above 25~GeV, making them unsuitable
for detecting the neutrino flux from light sneutrinos. 

The capture rate for DM particles in the core of the Sun depends on the DM--nucleus scattering 
cross-section, as well as on the DM velocity distribution and local density. After being captured, 
the DM annihilates into Standard Model particles, which further decay into neutrinos that can be 
observed at the Earth. The capture rate is approximated as ~\cite{Jungman:1995df,Gould:1987ju}
\begin{eqnarray*}
C_{\lsp} &=& 4.8\times 10^{24}\,{\rm s}^{-1}  
               \left(\frac{\rho_{\lsp}}{0.3\,{\rm GeV/cm^3}}\right) 
               \left(\frac{270\, {\rm km/s}}{\bar v}\right) \nonumber \\
   & & \times  \displaystyle\sum_{i} \left( \frac{\sigma^{\rm SI}_{{\lsp} i}}{10^{-40}\,{\rm cm}^2}\right) 
                 \frac{ f_i \phi_i}{m_{\lsp} m_{N_i}} F_i(m_{\lsp}) S(m_{\lsp}/m_{N_i})
\end{eqnarray*}
where  $m_{N_i}$, the mass of the nuclear species $i$, and $m_{\lsp}$ are given in GeV. 
$\bar{v}$ is the DM velocity dispersion, $f_i$ is the mass fraction of element $i$ in the Sun, and $\phi_i$ its distribution. 
$F_i(m)$ is a form factor suppression and $S$ a kinetic suppression factor. 
For these parameters, we use the values listed in Tables~8 and 9 of Ref.~\cite{Jungman:1995df}. 
Finally, $\sigma^{\rm SI}_{\lsp i}$ is the elastic scattering cross-section on point-like nucleus. 
 
In models where the DM is not self-conjugate, one can get different capture rates  for particles and 
antiparticles. Furthermore both particle--particle (antiparticle--antiparticle) and particle--antiparticle 
annihilation channels exist, in our case $\lsp\lsp \rightarrow \nu\nu$ ($A_{\chi\chi}$) and 
$\lsp\lsp^* \rightarrow X\bar{X}$ ($A_{\chi\bar\chi}$).
The equations describing the evolution of the number of DM (anti-)particles, $N_\chi(N_{\bar\chi})$, 
are then 
\begin{eqnarray}
   \dot{N}_\chi& =& C_\chi - 2A_{\chi\chi} N_\chi^2-A_{\chi\bar\chi} N_\chi N_{\bar\chi}\,, \nonumber\\
   \dot{N}_{\bar\chi}& =& C_{\bar\chi} - 2 A_{\bar\chi\bar\chi} N_{\bar\chi}^2-A_{\chi\bar\chi} N_\chi N_{\bar\chi} \,,
\label{eq:ndot}
\end{eqnarray}
The annihilation rates can be approximated as
\begin{equation}
A_{\chi\chi(\chi\bar\chi)}=\frac{\langle \sigma v\rangle_{\chi\chi(\chi\bar\chi)}}{V_{eff}} \,,
\end{equation}
where $V_{eff}= 5.8\times 10^{30} {\rm cm}^3 (m_\chi/{\rm GeV})^{-3/2} $  is the effective volume of the core of the 
Sun~\cite{Griest:1986yu}.
If the capture and annihilation rates are sufficiently large, equilibrium is reached  and the annihilation rate 
is only determined by the capture rate. We assume this to be the case. To take into account the different 
capture and annihilation rates for particles and antiparticles, we define
\begin{equation}
   \beta= \frac{C_\chi}{C_{\bar\chi}} \,, \quad
   \alpha=\frac{A_{\chi\chi}}{A_{\chi\bar\chi}} = \frac{A_{\bar\chi\bar\chi}}{A_{\chi\bar\chi}} \,, \quad
   x=\frac{N_\chi}{N_{\bar\chi}} \,.
\end{equation}
Then, after equilibrium is reached, $\dot{N}_\chi=\dot{N}_{\bar\chi}=0$ and we can solve Eq.~\ref{eq:ndot}. 
The annihilation rate at present is determined by the  capture rates as well as $\alpha$:
\begin{eqnarray}
\Gamma_{\chi\bar\chi} & =& A_{\chi\bar\chi} N_\chi N_{\bar\chi} =\frac{C_{\bar\chi}}{1+\alpha x}\,, \\\nonumber
\Gamma_{\chi\chi} & =& A_{\chi\chi} N_\chi N_{\chi} = \frac{1}{2}C_\chi\left(\beta -\frac{1}{1+\alpha x}\right)\,,\\\nonumber
\Gamma_{\bar\chi\bar\chi} & =& A_{\bar\chi\bar\chi} N_{\bar\chi} N_{\bar\chi} =\frac{1}{2}C_{\bar\chi}\frac{\alpha x}{1+\alpha x}  \,,
\end{eqnarray}
where 
\begin{equation}
x=\frac{1}{2\alpha\beta} \left[1-\beta + \left((\beta-1)^2+4\alpha^3\beta\right)^{1/2}\right] \,.
\end{equation}
The total neutrino spectrum at the Earth, assuming self-annihilation channels are solely into neutrino pairs,
is given by 
\begin{eqnarray}
  \frac{d\phi_\nu}{dE_\nu} &=& \frac{1}{4\pi d^2} \left(
       \Gamma_{\chi\chi} Br_{\nu\nu}\, \frac{dN_{\nu\nu}}{dE} +
       \Gamma_{\chi\bar\chi} \sum_f Br_{f\bar{f}}\, \frac{dN_f}{dE} \right)\,, \nonumber\\
  \frac{d\phi_{\bar\nu}}{dE_{\bar\nu}} &=& \frac{1}{4\pi d^2} \left( 
       \Gamma_{\bar\chi\bar\chi} Br_{\bar\nu\bar\nu} \, \frac{dN_{\bar\nu\bar\nu}}{dE} + 
       \Gamma_{\chi\bar\chi} \sum_f Br_{f\bar{f}}\, \frac{dN_f}{dE} \right) \,,
\end{eqnarray}
where $d=1.5\times 10^8$~km is the distance from the Sun to the Earth,
$Br_{\nu\nu}$ is the branching fraction for annihilation into neutrino pairs 
$Br_{f\bar{f}}$ the branching fraction into each particle/antiparticle final state $f\bar{f}$. 
$N_f$ and $N_{\nu\nu}(N_{\bar\nu\bar\nu})$ are the neutrino spectra resulting from 
those annihilations. Here  $dN_{\nu\nu}/dE$ is simply proportional to a delta function.
The neutrino spectrum originating from different annihilation channels into SM particles and 
taking into account oscillations was computed in~\cite{Cirelli:2005gh}, we use the tables given 
there. Note that for the neutrino pair an average over the three flavours in the annihilation process is assumed. 
For the one light sneutrino scenario this is not the case, however this is still a good
approximation since almost perfect 3-generation mixing is expected for  neutrinos below 
10~GeV propagating in the Sun~\cite{Cirelli:2005gh}. 
 
For light DM particles, the process of evaporation from the Sun can be important, this effect
was estimated in \cite{Gould:1987ju,Griest:1986yu}. We modify Eq.~\ref{eq:ndot} accordingly 
and solve it iteratively. We find that for the range of annihilation cross-sections of our scenarios, 
the evaporation affects significantly DM particles lighter than 3~GeV and is irrelevant 
for heavier DM particles.

Finally, to compare with the data, one must compute the muon flux for upward events~\cite{Erkoca:2009by}
\begin{equation}
   \frac{d\phi_\mu}{dE_\mu} =
\int\limits_0^{\mlsp} dE_{\nu}  \frac{d\phi_\nu}{dE_{\nu}}
  \int\limits_0^\infty dz \int\limits_0^{E_{\nu}} dE^{'}_{\mu}
\frac{dP_{cc}(E_{\nu},E^{'}_{\mu})}{dzdE_\mu^{'}}
P_{surv}(E_\mu^{'},E_\mu) \delta(E_{\mu}-E_{\mu}(E^{'}_{\mu},z))
\end{equation}
where the survival probability for a muon of energy $E'_\mu$ and final energy $E_\mu$ 
\begin{equation}
  P_{surv}(E_\mu^{'},E_\mu) = \left(\frac{E_\mu}{E_\mu^{'}}\right)^y 
                                              \left(\frac{\alpha+\beta
E_\mu^{'}}{\alpha+\beta E_\mu}\right)^y
\end{equation}
and the muon energy lost after propagating a distance $z$ is 
\begin{equation}
    E_{\mu}(E^{'}_\mu,z)=  e^{-\beta\rho z}
E^{'}_{\mu}-\frac{\alpha}{\beta}(1-e^{-\beta\rho z})
\end{equation}
Here $y=m_\mu /(c\tau\alpha\rho)$, $\tau$ is the muon lifetime,   
$\rho=2.6$~g/cm$^3$ the rock density and  $\alpha=2\times 10^{-3}$~GeV\,cm$^2$/g,
$\beta=3\times 10^{-6}$~cm$^2$/g characterize the average energy loss of the muon 
traveling through rock or water. $dP_{CC}$ is the probability for a neutrino with energy $E_\nu$ to be converted into a muon of energy $E_\mu$
over a distance $dz$ through charged current interactions.

The event rate takes into account the effective area of the detector. For Super-K, 
the detector is cylindrical with a radius $R=18.9$\,m and a height $36.2$\,m. The muons that 
are stopped in the first 7\,m are not observed, this corresponds to the muon energy threshold 
of $E=1.6$ GeV. Furthermore, only muons of energy larger than $7.7$~GeV go through the 
detector. This means that for our allowed scenarios with DM masses below 8 GeV, almost all 
the muons will be stopped within the detector.
We therefore compute the rate for stopped muons only. To do this we take into account the 
zenith angle $\theta_z$  when computing the effective area of the detector and average 
over $-1<\cos\theta<0$~\cite{belanger_prepa}.

\begin{figure}[t!]\centering
\includegraphics[width=14cm]{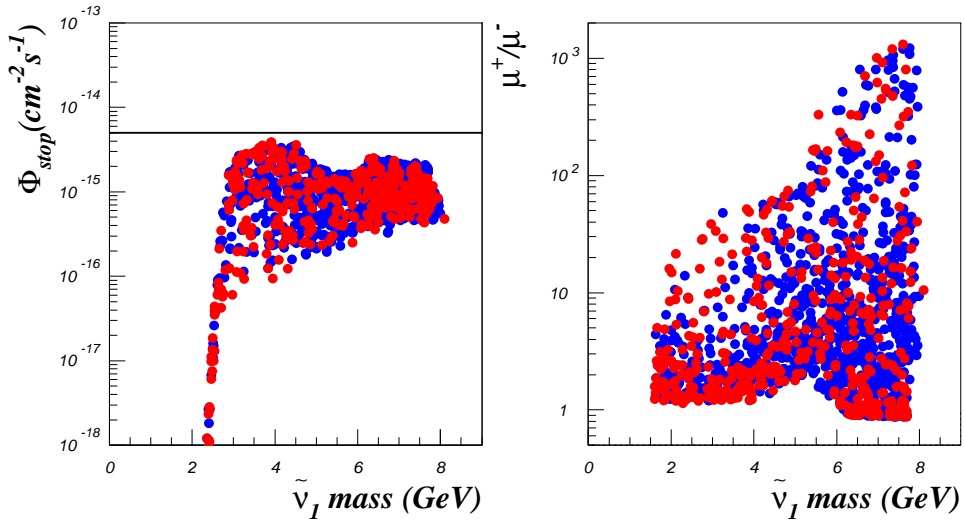}
\vspace*{-2mm}
\caption{Predicted flux from stopped muons from the Sun in the Super-K detector 
for $\bar v=270$~km/s (left) and the corresponding ratio of antimuon to muon fluxes (right), 
for the allowed scan points of Section~\ref{sec:onesnu}, cf.\ Figs.~\ref{fig:good1} and 
\ref{fig:good1dd}.
The red (blue) points have a relic density within (below) the $3\sigma$ WMAP range. 
The horizontal line in the lhs plot shows the extrapolated Super-K limit. 
\label{fig:nu}}
\end{figure}

In Fig.~\ref{fig:nu}, we display the total muon/antimuon rate as a function of the $\lsp$ mass 
for the scan results of Section~\ref{sec:onesnu} (only the points that pass all constraints are used). 
The Super-K limit as extrapolated
from~\cite{Niro:2009mw} is also displayed. Here we take  
$\bar v=270$~km/s for the DM velocity dispersion and $\rho_{\lsp}=\xi\times 0.3$~GeV/cm$^{3}$ 
with the same re-scaling factor $\xi$ as in Section~\ref{sec:onesnu} when the sneutrinos do 
not make up all the DM.  
Furthermore, we take the values of Eq.~(\ref{eq:scalar}) for the quark coefficients. The larger 
rates are found for sneutrinos lighter than $m_b$ that annihilate dominantly into $\tau^+\tau^-$ or 
neutrino pairs, since these modes give a harder neutrino spectrum. Some of our scenarios are within 10\% of  
the experimental limit in the mass range $3-4$~GeV.  In general, the expected rates  
are within one order of magnitude of the Super-K limit, except for
$m_{\lsp} <  3$~GeV where the evaporation effect becomes very important so that the 
neutrino flux is strongly suppressed. 
Note that these predictions depend on the DM velocity and average density. In particular a lower 
$\bar v$ would lead to larger neutrino rates. At the same time a lower DM velocity would relax the DD constraints. 

As already mentioned, one characteristic of our mixed sneutrino DM scenario is that the 
scattering rates on nucleons can differ significantly for particles and antiparticles. In fact, 
the capture of $\tilde\nu^*$ is in general more efficient than that of $\tilde\nu$ due to the destructive 
interference between the Z and Higgs exchanges in the latter case. As a result, the flux for 
antineutrinos and thus of antimuons is often much larger than that for muons, see the rhs plot 
in Fig.~\ref{fig:nu}. Distinguishing muon from antimuon events would therefore provide an additional 
test of this model.

\subsection{Photons}

As mentioned previously, when $m_{\lsp}\gsim 5$~GeV the dominant $\lsp$ pair annihilation 
channels are $\nu_\tau\nu_\tau$ or $b\bar{b}$, while for lighter DM the charged fermion channels are 
$c\bar{c}$ and  $\tau^+\tau^-$, with both channels having similar rates. 
Below ca.\ 1.5 GeV, the annihilation is purely into neutrinos. 
The annihilation channels into charged fermions leave a signature in photons, antiprotons and positrons. 
Photons are particularly interesting as the Fermi-LAT satellite is currently taking data in this channel.
(The signature in charged cosmic rays is discussed in the next subsection.) 
The Fermi-LAT collaboration 
has obtained its first limits on the flux of photons originating from DM annihilations.
In particular limits on $\sigma v$ were extracted from observations of dwarf galaxies~\cite{Abdo:2010ex}. 
The best one is obtained from Ursa Minor and corresponds to  
$\sigma v> 7\times 10^{-26}\,{\rm cm^3\,s^{-1}}$ for $m_{\rm DM}=10$~GeV 
assuming an annihilation entirely into $b\bar{b}$. 

\begin{figure}[t]\centering
\includegraphics[width=9cm]{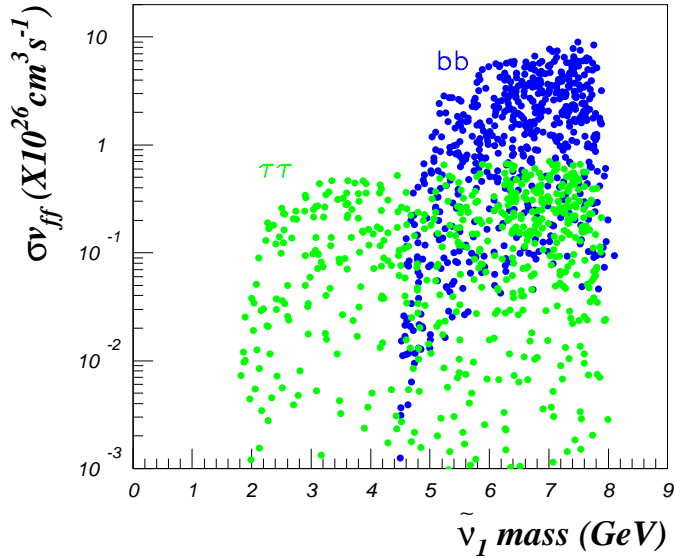}
\vspace*{-2mm}
\caption{Annihilation cross-sections $\sigma v_{f\bar{f}}$ for $f=b,\,\tau$ as a function of 
              $\mlsp$, for the allowed scenarios of Section~\ref{sec:onesnu}. 
\label{fig:gamma}}
\end{figure}

To study the sensitivity to light sneutrino DM, we compute 
 the annihilation cross-section into the $f\bar{f}$ channel, $\sigma v_{f\bar{f}}$,  for the parameter points of 
Section~\ref{sec:onesnu} that pass all constraints. 
To this aim we assume a NFW dark matter profile~\cite{Navarro:1995iw}.
The results  are shown in Fig.~\ref{fig:gamma} for $f=b$ and $\tau$. 
The $c\bar{c}$ channel gives similar results to $\tau^+\tau^-$. 

When $b\bar{b}$ is the dominant charged particle final state, $\sigma v_{b\bar{b}}$ ranges from  
$10^{-27}\,{\rm cm^3\,s^{-1}}$ to $10^{-25}\,{\rm cm^3\,s^{-1}}$. 
We therefore expect that some of our scenarios  
could be probed by Fermi-LAT once the analysis is extended to lower masses. 
When $\tau^+\tau^-$ is the dominant channel, $\sigma v_{\tau\bar\tau}$ can reach up to 
$10^{-26}\,{\rm cm^3\,s^{-1}}$. However, this channel gives a harder photon spectrum,
so that the sensitivity on the photon flux is expected to be almost an order of magnitude 
better than for the $b\bar{b}$ channel~\cite{Abdo:2010ex}. The Fermi-LAT satellite
could therefore also probe some of the scenarios with $\lsp$ masses below ca.\ 5 GeV. 
Note that for $m_{\lsp}\lsim 2$~GeV, the annihilation is dominated by neutrinos, leaving no signature 
in the photon channel. These scenarios are also the ones that because of evaporation have 
very suppressed rates in neutrino telescopes.

\subsection{Charged particles: positrons and antiprotons}

\begin{figure}[t]\centering
\includegraphics[width=7.4cm]{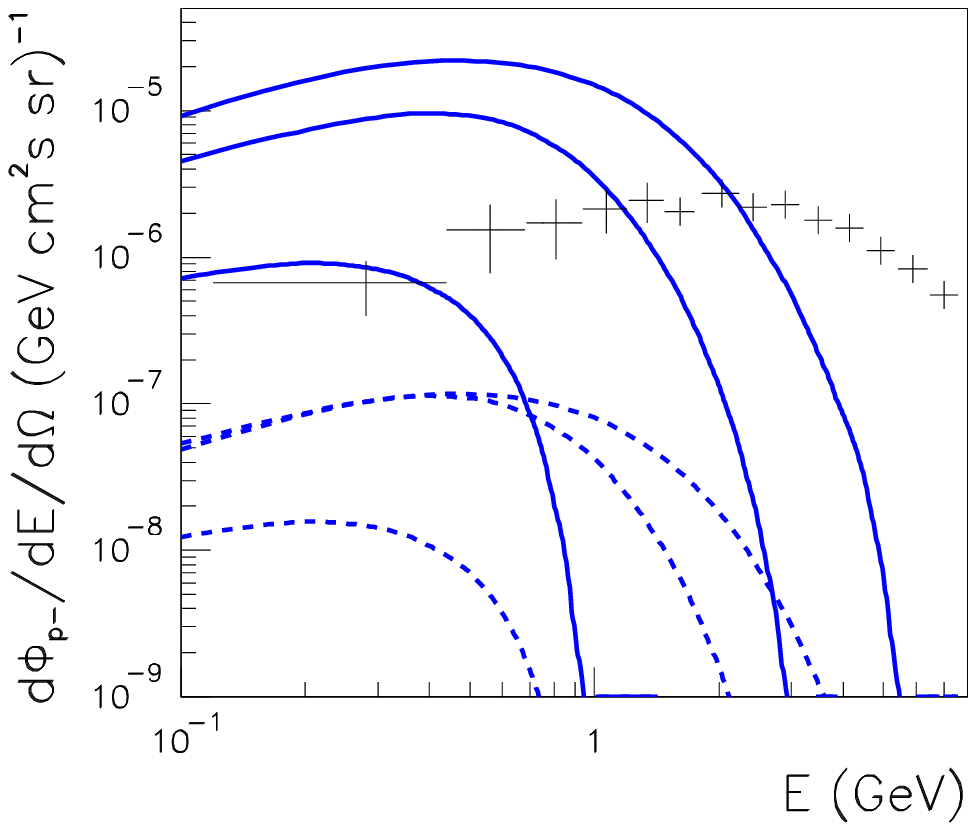}\quad
\includegraphics[width=7.4cm]{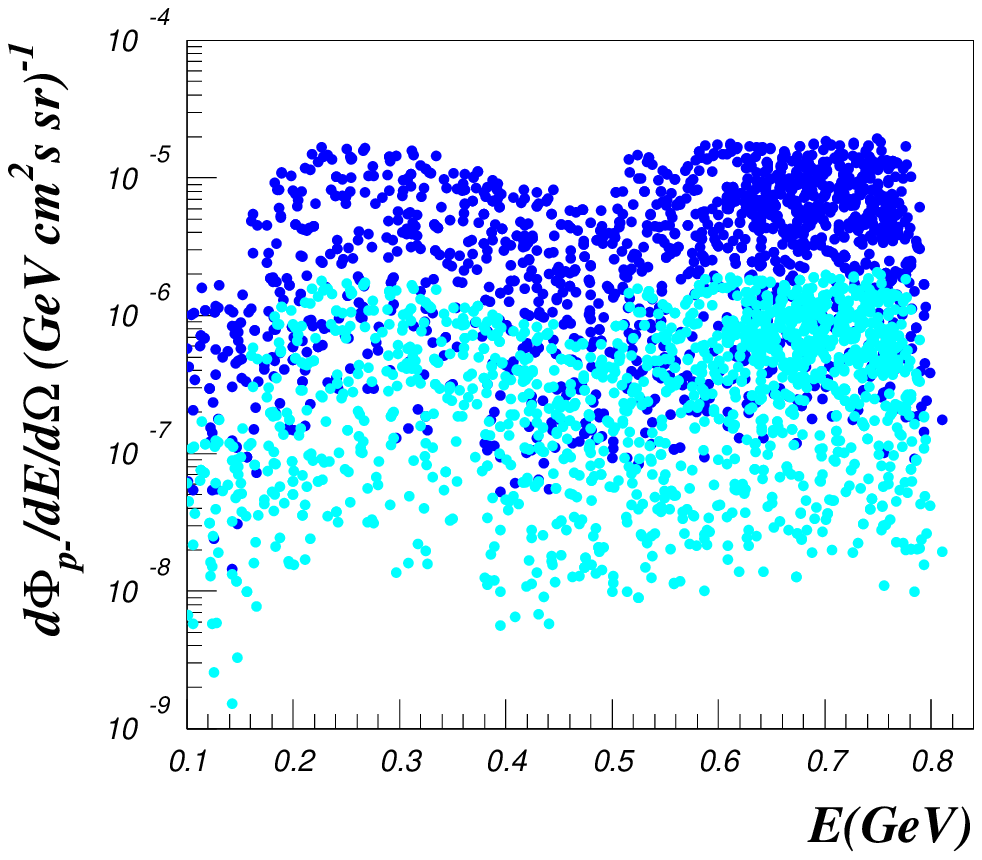}
\vspace*{-2mm}
\caption{On the left, the antiproton spectrum for 6 representative allowed points as explained in the text; 
the antiproton flux measured by PAMELA~\cite{Adriani:2010rc} is also displayed (crosses).
On the right, the differential flux for $\lsp$ annihilation into antiprotons at $E=0.1\mlsp$ including 
propagation for two sets of propagation parameters, MIN (light blue points) and MED (dark blue 
points), see Table~\ref{tab:prop}.
\label{fig:pbar} }
\end{figure}

\begin{figure}[t]\centering
\includegraphics[width=8cm]{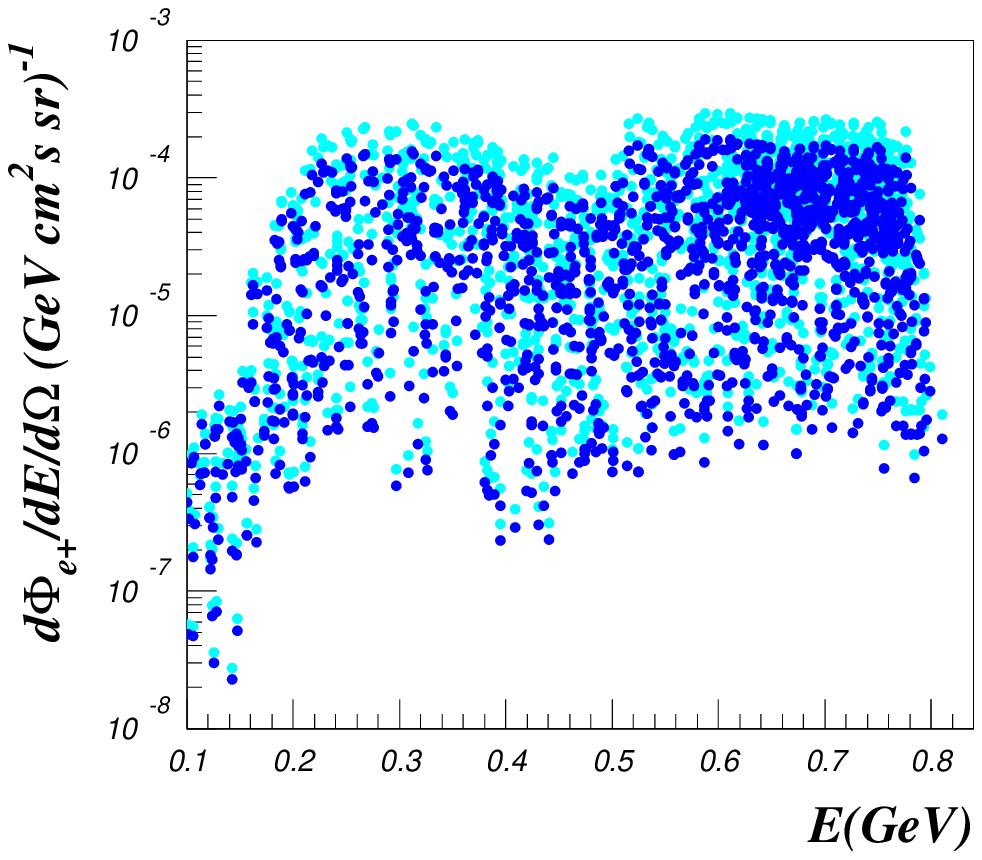}
\vspace*{-2mm}
\caption{Differential flux for $\lsp$ annihilation into positrons at $E=0.1\mlsp$ including propagation 
for two sets of propagation parameters, MIN (light blue points) and MED (dark blue points), 
see Table~\ref{tab:prop}.
\label{fig:posi} }
\end{figure}

The annihilation into charged particles will also leave a signature for antiprotons and positrons.  
The flux of antiprotons has been measured by PAMELA~\cite{Adriani:2010rc} together with the 
ratio of antiprotons to protons. 
It is well described by the expected background flux from standard astrophysical 
processes, the spallation of cosmic ray proton and helium nuclei over the interstellar medium. 
There are, however, large uncertainties in the theoretical predictions of the secondary fluxes,
due to uncertainties in the $\pbar$ production cross-sections, as well as in the parameters of 
the propagation models~\cite{Donato:2001ms}. 
Each of these uncertainties have been estimated to be around 25\% for the diffusion model 
with convection and re-acceleration described in~\cite{Donato:2001ms}.
Therefore there is still room for an additional contribution from DM annihilation, especially 
in the lower part of the spectrum where the experimental uncertainties are the largest. 
For an energy of $E_{\pbar}=0.56$~GeV, for instance, 
which roughly corresponds to the peak of the spectrum from a 5 GeV DM particle, the measured 
flux is $\Phi_{\bar p}=15.3^{+7.5}_{-3.7}\pm 0.9 \times 10^{-7}\;({\rm GeV\,cm^2\,s\,sr})^{-1}$. 
The left panel in Fig.~\ref{fig:pbar} shows antiproton spectra originating from $\lsp$ annihilation 
for $\mlsp = 1.4$, $3.9$ and $7.5$~GeV from left to right for either line-style. 
For each of these masses we have selected two scenarios corresponding to near maximal 
(full lines) and near minimal (dashed lines) flux. 
The selected scenarios are within 10\% of the absolute maximum/minimum fluxes.
As can be seen, the predictions can vary by around two orders of magnitude for a given mass, 
depending on the details of the parameter point. 
Indeed, while the antiproton fluxes from $\lsp$ annihilation alone can by far exceed 
the observed values in certain cases, they can also be more than one order 
of magnitude below the measurements and the expectations for the secondary spectrum. 
 
The spectra we just discussed were obtained using the MED set of diffusion parameters 
as given in Table~\ref{tab:prop}, together with a solar modulation in the force field approximation 
with $\phi_F=250$~MeV~\cite{Belanger:2010gh}. 
These diffusion parameters are in fact source of an important uncertainty in the flux. 
For illustration, the plot on the right in Fig.~\ref{fig:pbar} compares the antiproton flux at $E=0.1\mlsp$ 
obtained with the MED diffusion model (dark blue points) to that obtained with the MIN model 
(light blue points). The scattered points are the allowed scan points of the one-sneutrino case, 
and $E=0.1\mlsp$ was chosen because as the antiproton spectrum has its maximum at an energy 
that roughly corresponds to 10--15\% of the DM mass. 
The plot gives an estimate of the strong dependence of the primary flux on the propagation parameters
of nearly one order of magnitude just between the MIN and MED model. Similarly, it is possible to choose 
other propagation models (for example the MAX model) that further increase the flux by almost an order 
of magnitude. From Fig.~\ref{fig:pbar} we see that most of our scenarios are below the $1\sigma$ 
uncertainties of the measured flux when assuming the MIN propagation model, while strong constraints 
on our scenarios are expected assuming the MED propagation model.\footnote{A detailed fit to the 
antiprotons and $\pbar/p$ flux ratio will appear elsewhere.}

\begin{table}[htb]\centering
  \begin{tabular}{c|c|c|c|c}
  Model & $\delta$ & $K_0\;(\rm kpc^2/Myr)$ & $L\;(\rm kpc)$ & $V_C\;(\rm{km/s})$ \\\hline
  MIN & 0.85 & 0.0016 & 1 & 13.5 \\ \hline
  MED & 0.7 & 0.0112 & 4 & 12 \\ \hline
  MAX  & 0.46 & 0.0765 & 15 & 5 \\
  \end{tabular}
  \caption{Typical diffusion parameters that are compatible with the B/C 
  analysis~\cite{Maurin:2001sj,Donato:2003xg}. \label{tab:prop} }
\end{table} 

Let us now turn to the positron fluxes. These are displayed in Fig.~\ref{fig:posi},  
again assuming the MIN and MED propagation models and $E=\mlsp/10$. 
Here the MIN model gives a flux from $\lsp$ annihilation that is roughly a factor 
$1.5$ higher than that obtained with the MED model. 
PAMELA has measured the positrons to electron ratio~\cite{Adriani:2008zr} but has not yet 
released the positron fluxes. We can compare however with the secondary flux computed 
in~\cite{Delahaye:2008ua} which lies around 
$\Phi_{e^+} \approx 5\times 10^{-4}\,({\rm GeV\,cm^2\,s\,sr})^{-1}$ 
for the MED propagation model in the energy range $E=0.2-0.8$~GeV 
relevant for our scenarios; for the MIN model, the predictions can increase 
by roughly a factor 2. From Fig.~\ref{fig:posi} we see that the positron flux 
from $\lsp$ DM annihilation reaches at most   
$\Phi_{e^+} \approx 2\, (3) \times 10^{-4}\;({\rm GeV\,cm^2\,s\,sr})^{-1}$ in the MED (MIN) 
model and hence is always smaller than the secondary flux.

\section{Collider signatures of sneutrino DM}\label{sec:colliders}

\subsection{LHC}\label{sec:lhc}

An important characteristic of the light sneutrino DM scenario is the invisible decay of the light Higgs~\cite{ArkaniHamed:2000bq}.
At the LHC, the search for an invisible Higgs will be performed in the WW fusion channel 
with a signature in 2 tagged jets and missing $E_T$.
Defining the ratio $\zeta^2=\sigma(Hjj)/\sigma(hjj)_{SM} \times Br(h\rightarrow {\rm inv})$, 
the region to be probed with ${\cal L}=10$~fb$^{-1}$ at $\sqrt{s}=14$~TeV
corresponds to  $\zeta^2>0.38$ for a Higgs mass below 150~GeV~\cite{deRoeck}.
In our sneutrino DM scenarios, the light Higgs is SM-like and the invisible decay $h\to\lsp\lsp^*$
overwhelmingly dominant ($\approx 99\%$), making for a good Higgs 
discovery potential in the invisible channel.

The SUSY signatures also differ from the expectations in the conventional MSSM:
while squarks and gluinos have the usual cascade decays through charginos and 
neutralinos, with the same branching ratios as in the corresponding MSSM case 
(see, e.g., \cite{Drees:2004jm,Baer:2006rs}), here the 
charginos and neutralinos decay further into the $\lsp$ LSP. 
For 90\% of the allowed parameter points of our scans, both the $\neuto$ and the 
$\neutt$ decay to practically 100\% into $\nu\lsp$, leading to larger missing $E_T$ than 
naively expected for a LSP that weighs only a few GeV. (The $\neuto$ in fact always 
decays invisibly.)
Note also that in this case $\tilde q_R\to q\neuto$ and $\tilde q_L\to q\neutt$ give the 
same signature of jet+$E_T^{\rm miss}$, differing only in the jet-$p_T$ and $E_T^{\rm miss}$ 
distributions. 
The $\charg$ also decays directly into the LSP, with BR$(\charg\to l^\pm\lsp)\approx 100\%$ 
in the large majority of the cases.  
This means that decay chains involving charginos should on average have less missing 
$E_T$ than chains involving neutralinos. It also means that $\neutt \charg$ production 
leads to a single charged lepton rather than the usual trilepton signature. 
Furthermore if, as we have assumed, the LSP is the tau-sneutrino, 
the charged lepton will be a tau. 

This picture depends only little on most of the 
parameters that we have fixed in the scans, in particular the value of the higgsino mass 
parameter $\mu$ hardly influences the picture (lowering $\mu$ increases a bit the 
$\charg\to W^\pm\neuto$ decays, but these add to the single lepton events, although 
in a flavour-democratic way). 
The situation is, however, different for the first and second generation 
slepton masses: lowering them far enough can open up the 
$\neutt\to \ell^\pm \tilde\ell_L^\mp$ and/or $\neuto\to \ell^\pm \tilde\ell_R^\mp$ 
decay channels ($\ell=e,\,\mu$), and this crucially influences the 
experimental signatures. 
Setting, for instance, $m_{\widetilde L_{1,2}}=m_{\widetilde R_{1,2}}=100$~GeV in 
the DM allowed scan points, while assuming that the RH $\tilde\nu_{e,\mu}$ are heavy, 
gives $\neutt\to \ell^\pm \tilde\ell_L^\mp$ 
decay branching ratios of typically up to 30\%, and up to  50\% if the decay into 
$\nu_\tau\tilde\nu_{1\tau}$ is suppressed by a small mixing angle (the rest goes into 
LH $\tilde\nu_{e,\mu}$'s). Note  that decays into staus are mostly absent in this case. 
Taking, as may seem more natural, 
$m_{\widetilde L_1}=m_{\widetilde L_2}\approx m_{\widetilde L_3}$, 
$m_{\widetilde N_1}=m_{\widetilde N_2}\approx m_{\widetilde N_3}$ and 
$|A_{\tilde\nu_{e,\mu}}|\lsim |A_{\tilde\nu_\tau}|$ brings us back to the situation 
that invisible neutralino decays are overwhelmingly dominant.  
The $\charg$ decays democratically into all three lepton flavours 
in this case, $\charg\to l^\pm\tilde\nu_{1l}$, with only a small preference for $\tau^\pm\tilde\nu_{1\tau}$. 
Finally, for $A_{\tilde\nu_{e,\mu}}\to 0$ we recover the situation discussed in the previous 
paragraph, with $\neutt\to \nu_\tau\tilde\nu_{1\tau}$ and $\charg \to \tau^\pm\tilde\nu_{1\tau}$ 
having practically 100\% branching ratio.

A detailed study of the LHC potential to resolve the light sneutrino DM scenario, including 
in particular the determination of the DM mass from $\tilde q\to q'\charg\to q'l^\pm\tilde\nu_{1l}$ 
events, is left for future work.

\subsection{ILC signatures}

At an international linear collider (ILC) with $\sqrt{s}=500$~GeV, the main production for 
the light Higgs is $e^+e^-\to Zh$. This allows for precision measurements 
even if the $h$ decays entirely via invisible modes as is the case in our model~\cite{AguilarSaavedra:2001rg}. 
The main SUSY production processes at the ILC 
are $\charg\tilde\chi^\mp_1,\neuto\neutt,\neutt\neutt,\lsp\lsp^*$ 
as well as $\tilde\tau_1^+\tilde\tau_1^-$ 
(and maybe selectrons and smuons, depending on the scenario).

\begin{figure}[t]\centering
\includegraphics[width=8cm]{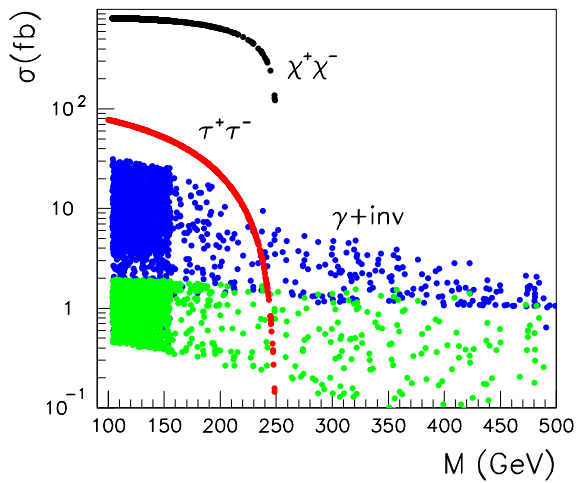}
\vspace*{-2mm}
\caption{Scatter plot of the unpolarized cross-sections of 
$e^+e^-\rightarrow \tilde\chi^\pm_1\tilde\chi^\mp_1$ (black) and 
$\gamma+{\rm inv}$ (green) at $\sqrt{s}=500$~GeV as a function of $\mchar$, and of 
$e^+e^-\rightarrow \stau_1^+\stau_1^-$ (red) as a function of $m_{\stau_1}$ for the 
allowed scenarios with one light sneutrino. Also shown is the single photon cross-section 
assuming $m_{\tilde{e}_R}=m_{\tilde{e}_L}=m_{\tilde{\tau}_R}$. 
\label{fig:lc} }
\end{figure}

Chargino-pair production has the largest cross-section and should be easily measurable, 
as is illustrated in Fig.~\ref{fig:lc}.  
The peculiarity of the sneutrino DM model is that the chargino decay gives a single charged 
lepton plus missing energy, see the discussion in the previous subsection. 
The cross-section for stau-pair production can reach 70\,fb if the staus are kinematically accessible.  
This occurs, however, only for a small fraction of the successful DM models, cf.~Fig.~\ref{fig:stau}. 
The rather heavy staus in our scenarios are a consequence of setting $m_{\widetilde L_3}=m_{\widetilde R_3}$.  
This assumption was not important for the DM study but has a strong impact here. Furthermore, 
the assumption of heavy selectrons and smuons means that the most favourable production
processes, notably $e^+e^-\to\tilde e^+\tilde e^-$ and $e^+e^-\to\tilde\mu^+\tilde\mu^-$, 
are kinematically not accessible.  
Relaxing this assumption, for example by assuming universality in the slepton masses, could give 
large selectron/smuon pair-production cross-sections and would also affect chargino-pair production,  
which depends on the mass of the LH $\tilde\nu_{e}$.  

As mentioned, neutral particles have large branching fractions into invisible states. In fact, 
the  $\neuto,\neutt\rightarrow \lsp \nu$ decays both have nearly 100\% branching fractions. 
Therefore, $\lsp\lsp^*$ and $\tilde\chi^0_i\tilde\chi^0_j$ ($i,j=1,2$) pair production will all 
contribute to the single photon cross-section. 
We have computed the total single photon cross-section exactly using 
calcHEP~\cite{Pukhov:2004ca} with the cuts 
$p_T^\gamma>4$~GeV and  $\theta_{beam \gamma}>10^\circ$. 
Monte Carlo simulations of this process including backgrounds have shown that cross 
sections around $1.6$~fb could be detectable  using beam polarisation ~\cite{Baer:2001ia,Birkedal:2004xn}
\footnote{More detailed analyses including detector simulation  for the ILC 
were also performed for DM masses above 100~GeV~\cite{Bartels:2009fa,Murase:2010he}, it remains
to be seen how can this be applied to our model where particles of different masses contribute to the 
single photon channel.}.
For the allowed points from Section~\ref{sec:onesnu}, the predictions for  the unpolarised cross-section
are mostly below 1~fb although the cross-section can reach  up to 2~pb  when $M_2< 150$~GeV, cf.\ the green points in Fig.~\ref{fig:lc}. The dominant contribution arises from the $\lsp\lsp^*$ channel.
The mass of the selectron is again a crucial parameter for these processes. To 
illustrate its impact we have also computed the single photon cross-section fixing   
$m_{\tilde{e}_R}=m_{\tilde{e}_L}=m_{\tilde{\tau}_L}=m_{\tilde{\tau}_R}$ (instead of 1~TeV). 
This increases the single photon cross-section by up to  one order of magnitude, see the blue points 
in  Fig.~\ref{fig:lc}. This is due mainly to the increase in neutralino production.  

Note finally that although there can be  an important additional Higgs production mode
through the decay  $\snt\rightarrow \lsp h$, this is not detectable since the Higgs decays
invisibly. Furthermore, in our model the $\tilde\nu_2$ is often too heavy to be pair 
produced with $\sqrt{s} = 500$~GeV.

\section{Conclusions}\label{sec:conclusions}

In supersymmetric models with Dirac neutrino masses, a weak-scale trilinear $A_{\tilde\nu}$ 
term that is not proportional to the small neutrino Yukawa couplings can  induce a sizable 
mixing between LH and RH sneutrinos and render the lighter sneutrino mass 
eigenstate a viable dark matter candidate. In particular, the mixed sneutrino can be an 
excellent candidate for light SUSY DM with mass below $\sim$10~GeV, as we have shown 
in this paper.

To obtain a very light $\lsp$ requires some fine-tuning, as the soft-breaking terms in 
the sneutrino mass matrix are all of the weak scale. 
Moreover, a viable sneutrino DM candidate requires enough mixing 
to provide sufficient pair-annihilation, while on the other hand the mixing should not be too large 
in order not to exceed the DD limits or contribute too much to the  Z invisible decay width. 
In addition, the 1-loop diagrams involving sneutrinos induce a negative correction to the light 
Higgs mass, which can attain a few GeV for large $\anu$.
In a random scan over $10^5$ points for the case of one light sneutrino contributing to the DM, 
we found close to 2000 points that survive all present constraints, including the Z invisible decay 
width, the Higgs and SUSY mass limits, as well as dark matter constraints from the relic abundance 
and the direct detection experiments. These points populate the region $\mlsp\approx1-8$~GeV. 
We also found that for very light $\lsp$ below ca.\ 5~GeV, small $M_2\approx100-150$~GeV 
is preferred in order to enhance $\lsp\lsp\to\nu\nu$ annihilation through t-channel wino exchange. 
In the case of three degenerate light sneutrinos, the constrains are much stronger and 
only few points pass all constraints. 

Our results for direct and indirect detection of light sneutrino DM can be summarized as follows.
First, the cross-section for the spin-independent elastic scattering on nuclei is predicted to 
be $\sigma^{\rm SI}>10^{-5}$~pb. That is at most within an order of magnitude of present 
limits for DM masses around 5--10~GeV, and it includes the region favoured by CoGeNT (and 
perhaps CRESST).
Second, sneutrino pair annihilation into $b\bar{b}$ and $\tau^+\tau^-$ can lead to large 
distortions of the positron and antiproton spectra at low energies. These are being probed 
by PAMELA.  Furthermore, the photon flux can be in the range being probed by FermiLAT,  
provided again the $\lsp$'s annihilate significantly into $b\bar{b}$ and $\tau^+\tau^-$.  
In all cases the signals are expected in the low energy range (ca.\ $0.1-3$~GeV), 
a region where measurements are not as precise as for higher energies.
Third, the neutrino flux resulting from $\lsp$ capture in the Sun can also be large. 
However, because the energy of the neutrinos is bounded by the DM mass, these scenarios 
are harder to probe than those with typical weak scale DM.
In fact the neutrino energy is below the threshold of the large detectors. 
Only SuperKamiokande has a low enough threshold to have some sensitivity to neutrinos 
from light sneutrino annihilation, though these neutrinos do not have enough energy to induce 
muons that go  through the detector. One can therefore use only events where muons are 
stopped in the detector. It is also interesting to note that the predicted flux for 
antineutrinos, and thus for antimuons, is often much larger than that for muons.   

While the scenarios with $\lsp$ DM that annihilates preferentially into $b{\bar b}$ have 
good prospects of being detected in the near future, those that  have a DM mass below 
ca.\ 2~GeV will mostly escape detection.  
Indeed direct detection experiments lack sensitivity for these masses.   
Moreover, the main annihilation channel in this case is into neutrinos, leaving low chances  
for indirect detection in photons, antiprotons and positrons; some possibilities remain if the 
annihilation into $\tau^+\tau^-$ is significant.  
Even the neutrino telescopes cannot make use of the large branching fraction for annihilation 
into neutrinos, because the evaporation process in the Sun strongly suppresses the 
neutrino flux for such light $\lsp$'s. 

Finally a mixed sneutrino LSP leaves distinct signatures in collider experiments.  
Most notably the light Higgs boson and the two lightest neutralinos decay almost exclusively 
into invisible modes, while decays of the lighter chargino give a single charged lepton 
plus missing energy. At the LHC, the typical cascade decays therefore are 
$\tilde q\to q\charg\to q'l^\pm\lsp$, $\tilde q_L\to q\neutt\to q\nu\lsp$ and 
$\tilde q_R\to q\neuto\to q\nu\lsp$, all giving different amount of missing $E_T$.
Moreover, $\charg\neutt$ production gives only a single charged lepton. 
At the ILC, $\lsp\lsp^*$ and $\tilde\chi^0_i\tilde\chi^0_j$ ($i,j=1,2$) pair production 
will all contribute to the single photon cross-section.

The details of the DM signatures at colliders depend of course on the assumptions that 
are made on the rest of the spectrum, in particular the first and second generation of
sleptons. The predictions in astroparticle experiments have uncertainties from astrophysical 
and nuclear parameters. These include uncertainties in the quark content of the nucleon, on
the local dark matter density and velocity distribution, on the dark matter halo profile  and,  
for charged cosmic rays, on the parameters of the  propagation model. 

Last but not least 
if signals are found in astroparticle and collider experiments, the challenge will be to 
determine the precise DM properties and the underlying new physics.
This requires in particular collider measurements of the masses and couplings of the DM 
and other new particles associated with it in order to refine and test the theoretical predictions 
for astroparticle observables. Moreover, if an $E_T^{\rm miss}$ signal is seen at the LHC, 
it should be confirmed in direct DM detection and vice versa, the agreement of the DM mass 
and cross-section determined in the two ways being a crucial test. 
To this end it will be interesting to investigate how well our light sneutrino DM scenarios 
can be resolved at the LHC by exploiting the $\tilde q\to q\charg\to q'l^\pm\lsp$ 
cascade decay. Such a signature has been suggested in~\cite{Polesello:2009rn} 
as a  method  of measuring mass differences in the MSSM. 
This will be a future work.

\section{Acknowledgments}
We thank M.~Cirelli, J.~Collar, E. Nuss  and T.~Schwetz for useful discussions.
EKP acknowledges the hospitality of LAPTH, LPSC and CERN, where most of this work was 
performed. This work is supported by HEPTOOLS under contract MRTN-CT-2006-035505.
This work is also supported in part by the GDRI-ACPP of CNRS and by the French 
ANR project {\tt ToolsDMColl}, BLAN07-2-194882.
The work of AP is supported by the Russian foundation for Basic Research, 
grant RFBR-08-02-00856-a, RFBR-08-02-92499-a and RFBR-10-02-01443-a.

\appendix\label{sec:app}

\section{Vertices involving sneutrinos}
The Feynman rules of relevant $\tilde{\nu}_1$ couplings include
\begin{eqnarray}
    Z^\mu \tilde{\nu}_1^*(p') \tilde{\nu}_1(p): && 
        - i \frac{e}{\sin 2 \theta_W} (p + p')^\mu  \sn^2  \, , 
    \nonumber \\
    h \tilde{\nu}_2^* \tilde{\nu}_1: &&
        - i e m_Z \frac{\sin(\alpha + \beta)}{\sin 2 \theta_W} 
    \cn \sn 
    - i \frac{1}{\sqrt{2}} A_{\tilde{\nu}} \cos \alpha 
    (\cos^2 \theta_{\tilde\nu}  - \sin^2\theta_{\tilde\nu} )\, ,
    \nonumber \\
    h \tilde{\nu}_1^* \tilde{\nu}_1: &&
   i e m_Z \frac{\sin(\alpha + \beta)}{\sin 2 \theta_W} \sin^2\theta_{\tilde\nu}  
    + i \sqrt{2} A_{\tilde{\nu}} \cos \alpha \cn \sn \, .
\end{eqnarray}

\begin{eqnarray}
W^\mu\tilde\nu_1\tilde{l_i}  \;\;&:& \;\;   i\frac{g}{\sqrt{2}} Z^l_{i1}  (p+p')^\mu \sn\nonumber\\
\tilde\chi^0_{i}\tilde\nu^*_1\nu  \;\;&:& \;\;    -i\frac{g}{2\sqrt{2}\sin 2\theta_W} (\cw N_{i2}-\sw N_{i1})  \sn (1+\gamma_5)\nonumber\\
\tilde\chi^+_{i}\tilde\nu^*_1 l \;\;&:& \;\;   -i\frac{g}{4m_W\cos\beta} \left[2m_W \cos\beta V_{1i} (1-\gamma_5) -
\sqrt{2} M_l U_{2i} (1+\gamma_5)\right] \sn \nonumber\\
\end{eqnarray}

Here $Z^l$ is the charged lepton mixing matrix that is diagonal in flavour space,
$N$ is the neutralino mixing matrix and $U,V$ the chargino mixing matrices. 
We use the SLHA notation for the MSSM part of the Lagrangian~\cite{Skands:2003cj}.

\providecommand{\href}[2]{#2}\begingroup\raggedright\endgroup

\end{document}